\documentclass[final]{IEEEtran}

\usepackage{lineno,hyperref}
\usepackage{comment}
\modulolinenumbers[5]
\usepackage[latin1]{inputenc}
\usepackage{graphicx}
\usepackage{amssymb,amsmath,amsfonts,amsthm}
\usepackage{algorithm,algorithmic}
\usepackage{subfigure}
\usepackage{xspace}
\usepackage[usenames,dvipsnames]{color}
\usepackage{verbatim}
\usepackage{url}
\usepackage{bm,dsfont}
\usepackage[T1]{fontenc} 
\usepackage{soul}
\usepackage{hyperref}
\hypersetup{
     colorlinks=true,
     linkcolor=cyan,
     filecolor=cyan,
     citecolor = cyan,    
     urlcolor=cyan,
     }
\usepackage{xcolor}
\usepackage{tabu}
\usepackage{multirow}
\usepackage{booktabs}
\usepackage{cite}
\hypersetup{
     colorlinks = true,
     linkcolor = [rgb]{0,0,0},
     anchorcolor = [rgb]{0,0,0},
     citecolor = [rgb]{0,0,0},
     filecolor = [rgb]{0,0,0},
     pagecolor = [rgb]{1,1,0},
     urlcolor = [rgb]{0,0,0},
     bookmarks,
        bookmarksopen = true,
        bookmarksnumbered = true,
        breaklinks = true,
        linktocpage,
        pagebackref,
        colorlinks = true,
        linkcolor = [rgb]{0,0,1},
        urlcolor  = [rgb]{0,0,1},
        citecolor = Orange,
        anchorcolor = [rgb]{0,0,0},
        hyperindex = true,
        hyperfigures
}

\newenvironment{list4}{
  \begin{list}{$\bullet$}{%
      \setlength{\itemsep}{0.05cm}
      \setlength{\labelsep}{0.2cm}
      \setlength{\labelwidth}{0.3cm}
      \setlength{\parsep}{0in} 
      \setlength{\parskip}{0in}
      \setlength{\topsep}{0in} 
      \setlength{\partopsep}{0in}
      \setlength{\leftmargin}{0.17in}}}
      {\end{list}}

\begin{document}
\title{\fontsize{0.83cm}{0.96cm}\selectfont A Survey on Reinforcement Learning-Aided\\ Caching in Mobile Edge Networks}
\author{
    \IEEEauthorblockN{Nikolaos Nomikos,~\IEEEmembership{Senior Member,~IEEE}, Spyros Zoupanos,\\ Themistoklis Charalambous,~\IEEEmembership{Senior Member,~IEEE}, Ioannis Krikidis,~\IEEEmembership{Fellow,~IEEE}\\ and Athina Petropulu,~\IEEEmembership{Fellow,~IEEE} 
    \thanks{\IEEEauthorblockA{N. Nomikos and I. Krikidis are with the IRIDA Research Centre for Communication Technologies, Department of Electrical and Computer Engineering, University of Cyprus, 1678 Nicosia, Cyprus. {\it Emails: {\tt surname.name@ucy.ac.cy}.}
    }}
    \thanks{\IEEEauthorblockA{S. Zoupanos is with the Department of Informatics, Ionian University, Corfu, Greece. {\it Email:} {\tt spyros@zoupanos.net}.
    }}
    \thanks{\IEEEauthorblockA{T. Charalambous is with the Department of Electrical Engineering and Automation, School of Electrical Engineering, Aalto University, Espoo, Finland. {\it Email:} {\tt themistoklis.charalambous@aalto.fi}.
    }}
    \thanks{\IEEEauthorblockA{A. Petropulu is with the Department of Electrical and Computer Engineering, Rutgers University, Piscataway, NJ, USA. {\it Email:} {\tt athinap@rutgers.edu}.
    }}
}}
\maketitle
\begin{abstract}
Mobile networks are experiencing tremendous increase in data volume and user density. An efficient technique to alleviate this issue is to bring the data closer to the users by exploiting the caches of edge network nodes, such as fixed or mobile access points and even user devices. Meanwhile, the fusion of machine learning and wireless networks offers a viable way for network optimization as opposed to traditional optimization approaches which incur high complexity, or fail to provide optimal solutions. Among the various machine learning categories, reinforcement learning operates in an online and autonomous manner without relying on large sets of historical data for training. In this survey, reinforcement learning-aided mobile edge caching is presented, aiming at highlighting the achieved network gains over conventional caching approaches. Taking into account the heterogeneity of sixth generation (6G) networks in various wireless settings, such as fixed, vehicular and flying networks, learning-aided edge caching is presented, departing from traditional architectures. Furthermore, a categorization according to the desirable performance metric, such as spectral, energy and caching efficiency, average delay, and backhaul and fronthaul offloading is provided. Finally, several open issues are discussed, targeting to stimulate further interest in this important research field.  \end{abstract}

\section{Introduction}\label{intro}
Currently, the wide commercial roll-out of fifth generation (5G) networks is increasingly becoming a reality, providing enhanced mobile broadband (eMBB) services to users, supporting ultra-reliable and ultra-low latency (URLLC) traffic for critical applications, as well as massive machine type communications (mMTC) for a broad range of Internet-of-Things (IoT) applications. As we move forward, the sixth generation (6G) vision is expected to materialize around 2030 and at that time, the International Telecommunication Union (ITU) predicts that the mobile data traffic volume will surpass 5 ZB per month, a 670-fold increase from 2010 \cite{itu2015}. Meanwhile, mobile subscriptions will more than triple, reaching 17.1 billion, as compared to 5.32 billion in 2010.  

Such figures necessitate novel wireless network design approaches and the adoption of breakthroughs from the field of machine learning (ML), offering promising results. A technique facilitating the progress of 6G communications is mobile edge computing (MEC) and caching, where computation-intensive tasks take place near the place of data collection and popular contents are in close proximity to users \cite{barbarossa2014spm, shi2016computer,liu2016cm}. In this way, centralized cloud-based computation is avoided, while the backhaul and fronthaul links are relieved from constant data fetching from remote web servers. {This setup reduces the computational and communication delays considerably, facilitating latency-intolerant applications that they were otherwise infeasible to provide.} At the same time, %\todo{traditional [meaning? centralized?]}
non-learning-based optimization techniques often fail to reach optimal solutions %\todo{[why? it needs some justification at this point; I make an attempt]}
{due to the dynamic nature of the problem that they fail to track}, while their complexity might be prohibitive for online network optimization. Thus, ML-aided MEC and caching can cope with the plethora of mobile data and answer the questions of where, when and what to cache, as well as which tasks should be computed at the edge \cite{zhang2019cst,luo2020fwc}. 

As the field of online ML-aided network optimization is of tremendous importance towards 6G evolution, this survey focuses on reinforcement learning (RL)-aided edge caching and provides a holistic approach by equally presenting the different cache-aided mobile edge architectures and the corresponding RL solutions.

\subsection{Mobile Edge Networks}\label{edge}
\begin{figure*}[t]
\centering
\includegraphics[width=0.75\textwidth]{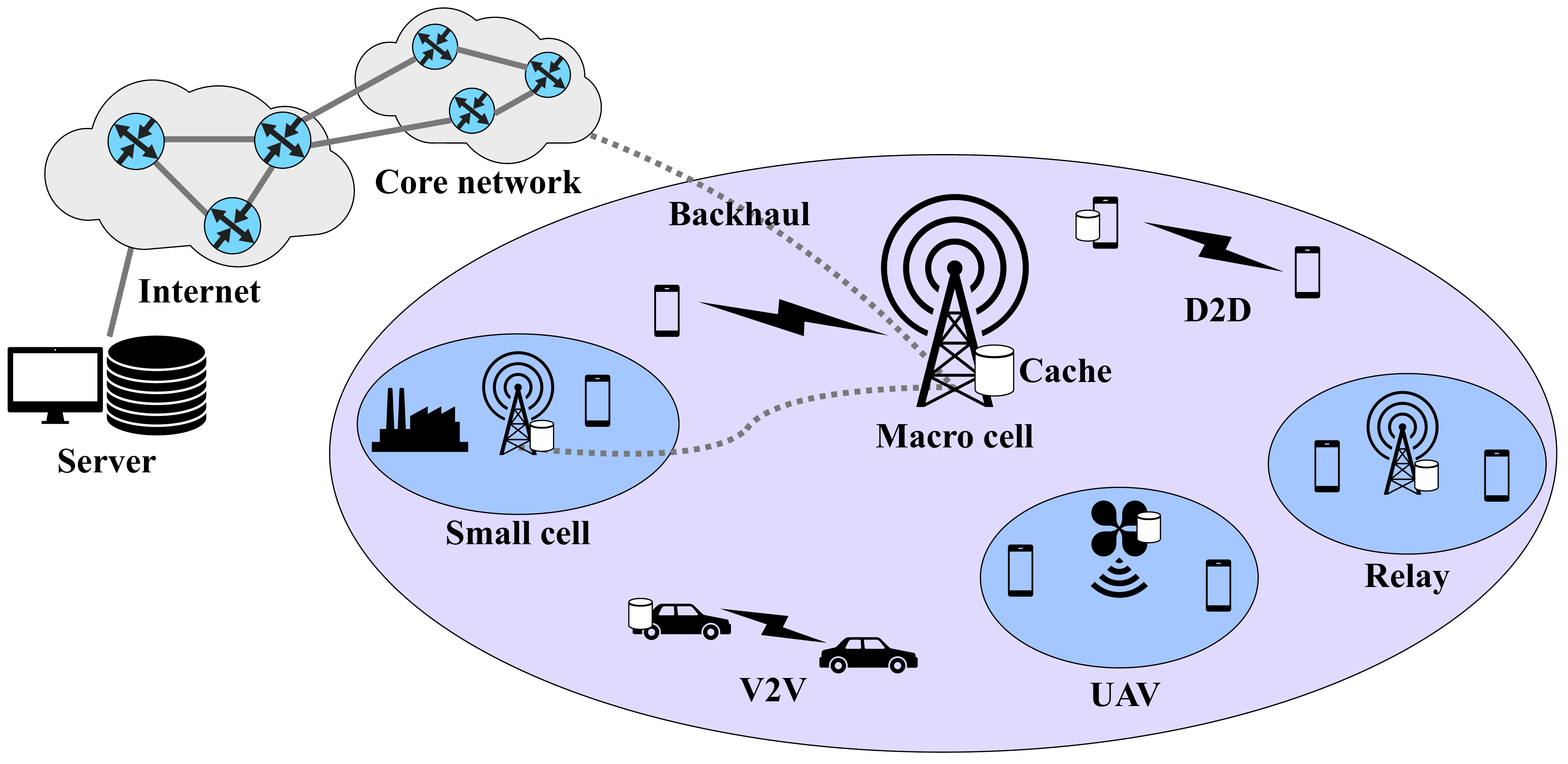}
\caption{A mobile wireless network with different cases of edge caching.}
\label{architecture}
\end{figure*}

Nowadays, mobile networks are struggling to satisfy the heterogeneous service requirements of coexisting users and devices. Conventional cellular architectures cannot provide throughput and delay guarantees and a radical departure is currently taking place, exploiting the cloud computing capabilities and the existence of edge network nodes \cite{pan2018iotj}. Cloud computing offers abundant processing power for various tasks, including among others, baseband processing for cloud radio-access networks (C-RANs), IoT applications with a massive number of sensors and mobile big data exploitation for optimizing network parameters \cite{wubben2014spm,zhou2018cm}. Unfortunately, centralized cloud architectures result in increased backhaul usage and end-to-end latency that might be intolerable for critical and real-time applications. So, bringing the computation resources closer to the network edge through the MEC paradigm has been proposed, as a remedy to the shortcomings of cloud computing \cite{shi2016computer,tran2017cm}. Also, the caching capabilities of edge nodes allow the contents to in proximity to the users, paving the way for performance gains, in the sense of latency minimization, throughput maximization, backhaul offloading, reduced operational expenditure due to energy savings, and finally, extended lifetime for the mobile terminals and the IoT devices \cite{liu2016cm}.

The MEC architecture requires from edge nodes to be equipped with storage capabilities for caching popular contents and avoiding constant fetching from remote web servers. Meanwhile, the wide range of different edge node types constitutes a challenging environment for optimizing the network performance \cite{piao2019iotj}. First, fixed cellular networks based on different tiers can benefit by storing content at small base stations, considering parameters, such as user mobility and interference among cells of the same of different tiers. Moreover, novel cellular architectures, comprising fog radio-access networks (F-RANs) rely on low-complexity fog access points (F-APs) instead of conventional BSs, where only a part of baseband processing takes place at the F-APs \cite{zhang2017wcm}. 

However, caching schemes should also aim at alleviating the fronthaul capacity constraints and improve the F-RAN performance. In addition, further extensions to traditional cellular networks are offered through cooperative communication and caching. More specifically, intelligent caching schemes can exploit the caching resources at different nodes, distributing the content for increased efficiency and robustness when specific nodes experience outages due to fading. In this context, cooperation between users through device-to-device (D2D) communication improves physical-layer aspects, such as coverage and transmit diversity while using storage at the devices for offloading the BSs \cite{wu2018wcm}. Finally, the introduction of highly mobile network nodes, providing wireless access and storage with increased flexibility in advanced use cases represents another fertile field of application for employing edge caching schemes. These mobile BSs include ground nodes, communicating in vehicle-to-vehicle (V2V) and vehicle-to-everything (V2X) scenarios, as well as unmanned aerial vehicles (UAVs), providing fast recovery after disasters and emergency situations, on-demand capacity provisioning and coverage in remote and rural areas \cite{gurugopinath2020network,zhao2019vtm}.  

An illustrative edge architecture is depicted in Fig.~\ref{architecture}, showing different cases of edge caching. More specifically, within the coverage area of a macro cell, a small cell caches content and serves users, requiring reliable and high throughput access, while a relay caches content that was transmitted from nearby users in the uplink and the macro BS in the downlink. Meanwhile, UAVs provide coverage to remote areas with limited coverage and store content that is scheduled for transmission towards the macro BS at a later moment, in order to reach the core network. Furthermore, various ad hoc communication paradigms exist, including cache-enabled devices communicating with each other, adopting D2D cooperation, as well as V2V communication in highly mobile environments.

\subsection{Machine Learning}\label{ml}

Research on employing machines to process large volumes of data, stemming from previously allocated tasks {or simulated scenarios}, towards learning to handle future tasks, has led to the tremendous growth of machine learning (ML) \cite{samuel1959ibm}. 
Machines exploit the wealth of data in various applications and interact with the environment in order to explore different actions and then, according to the observed reward, they adapt and exploit the actions yielding the highest reward for their next ventures. 
In mobile communication networks, a massive number of users and devices enjoy a broad range of services with different service requirements from heterogeneous network nodes, in terms of hardware capabilities. This explosive increase of wireless traffic requires highly complex network optimization solutions, posing difficulties to decision-making during resource allocation of bandwidth, power and storage. The adoption of online ML solutions in such challenging settings can lead to self-adaptive networks and accurate prediction of communication parameters, abiding to dynamic wireless conditions. In this way, network performance will be enhanced, offering improved Quality-of-Service (QoS) and resource efficiency. 

\begin{figure}[t]
\centering
\includegraphics[width=\columnwidth]{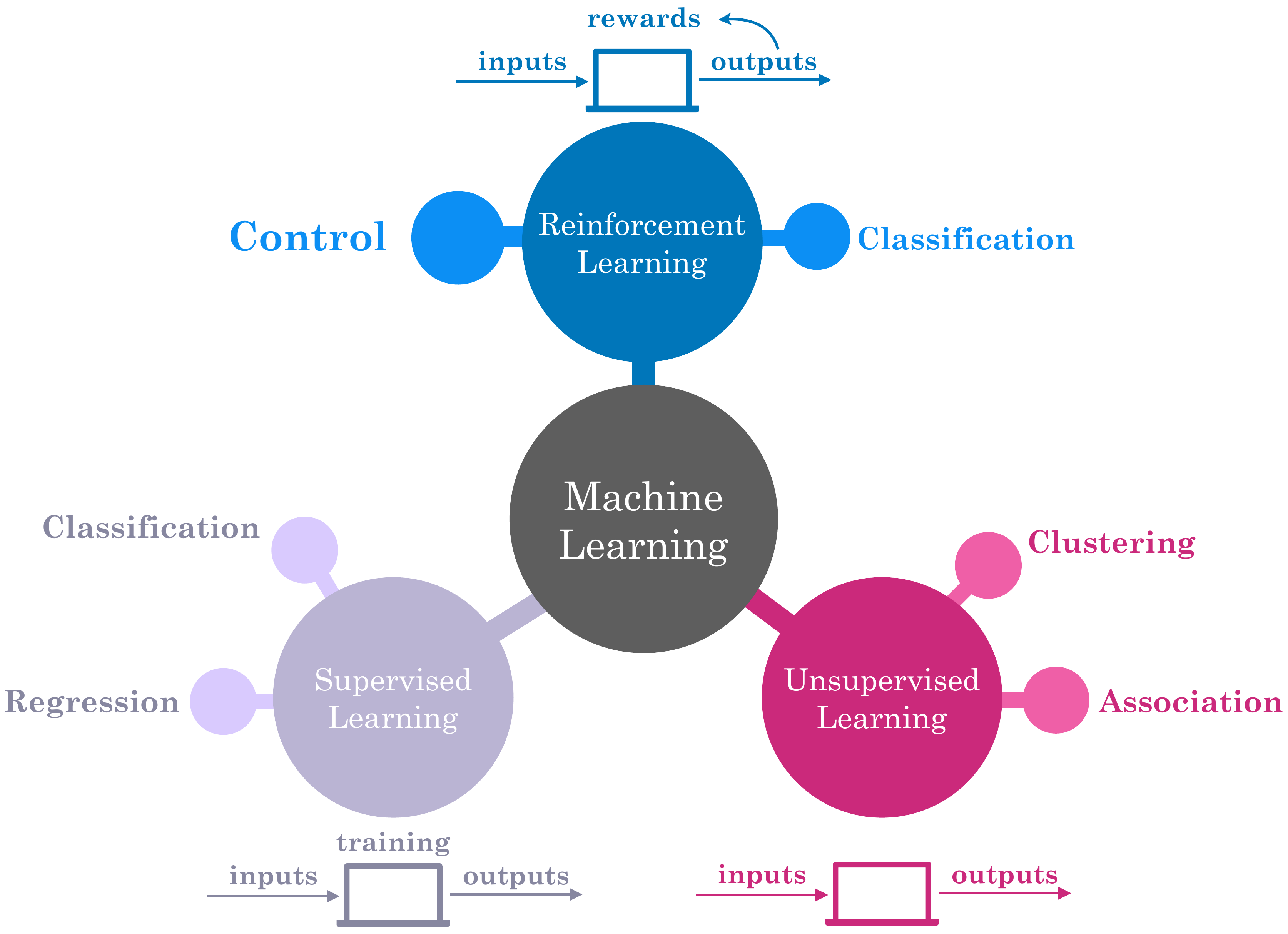}
\caption{Different classifications of ML.}
\label{ML_classification}
\end{figure}

ML is {mainly} classified into {three} different categories{; namely,} supervised learning, unsupervised learning, and reinforcement learning (RL) \cite{alpaydin2010mit}; see, Fig.~\ref{ML_classification}.
{In a finer categorization, one can find semi-supervised learning} and more recently, federated learning \cite{mcmahan2017aistats}. {In what follows, we discuss the different classes.}
\begin{list4}
\item \textbf{Supervised learning:} In supervised learning, the~algorithms rely on datasets, providing both the input and the output. Even though supervised learning provides improved decision making, the need for labeled data might be prohibitive in practice. Algorithms in this category include classification and regression analysis which can facilitate the characterization of data traffic and content popularity.
\item  \textbf{Unsupervised learning:} Unsupervised learning approaches rely on training data that do not include labeled output. Clustering is a popular method to develop unsupervised learning algorithms, enabling patter identification in datasets. In edge caching users can be cluster based on, for example, their desired contents, mobility, desire to cooperate with each other.
\item \textbf{Semi-supervised learning:} An intermediate approach regarding the nature of the available data has been followed with semi-supervised learning. In~this type of learning, both labeled and unlabeled data are exploited for the training.
\item \textbf{Reinforcement learning:} In RL, an agent's strategy is determined in an autonomous manner by considering the cost and reward of each action. Therefore, the~main idea of this type of learning is radically different as compared to the previous mentioned ones, which exploit historical data. Instead, RL algorithms are trained by using feedback on previously taken actions, adapting their behavior to the environment. In the edge caching case, various algorithms are used, such as Q-learning for predicting content request probability or deriving the popularity distribution. {While in supervised learning the model is trained with the correct answer, in RL there is no answer but the reinforcement agent makes the decision how to perform a given task. If there does not exist any training dataset, RL learns from its experience. Hence, unlike other approaches, RL is about taking suitable action to maximize a reward (e.g., best possible behavior or path) in a particular situation.}
\item \textbf{Deep learning:} Deep learning (DL) {is closely related to the above classes of ML. It} relies on multiple layers to form artificial neural network architectures for accurate decision making. In this hierarchical architecture lower-level features define higher-level ones, while feature extraction is autonomously performed. In edge caching cases, DL can provide near-optimal policies for content placement and pushing without excessive complexity, even though large volumes of training data should be available. An illustrative example of a DL architecture in the context of mobile edge networking is shown in Fig.~\ref{DL}. Here, the observation of the mobile edge environment leads to the formation of specific states that act as input to the deep neural network (DNN) for deciding the action that should be selected by the agent. Each action results in specific rewards, that in the long-term determine the efficiency of the DL policy. %\todo{[See also Fig.3 in https://ieeexplore.ieee.org/stamp/stamp.jsp?arnumber=8605302]}
\begin{figure}[t]
\centering
\includegraphics[width=\columnwidth]{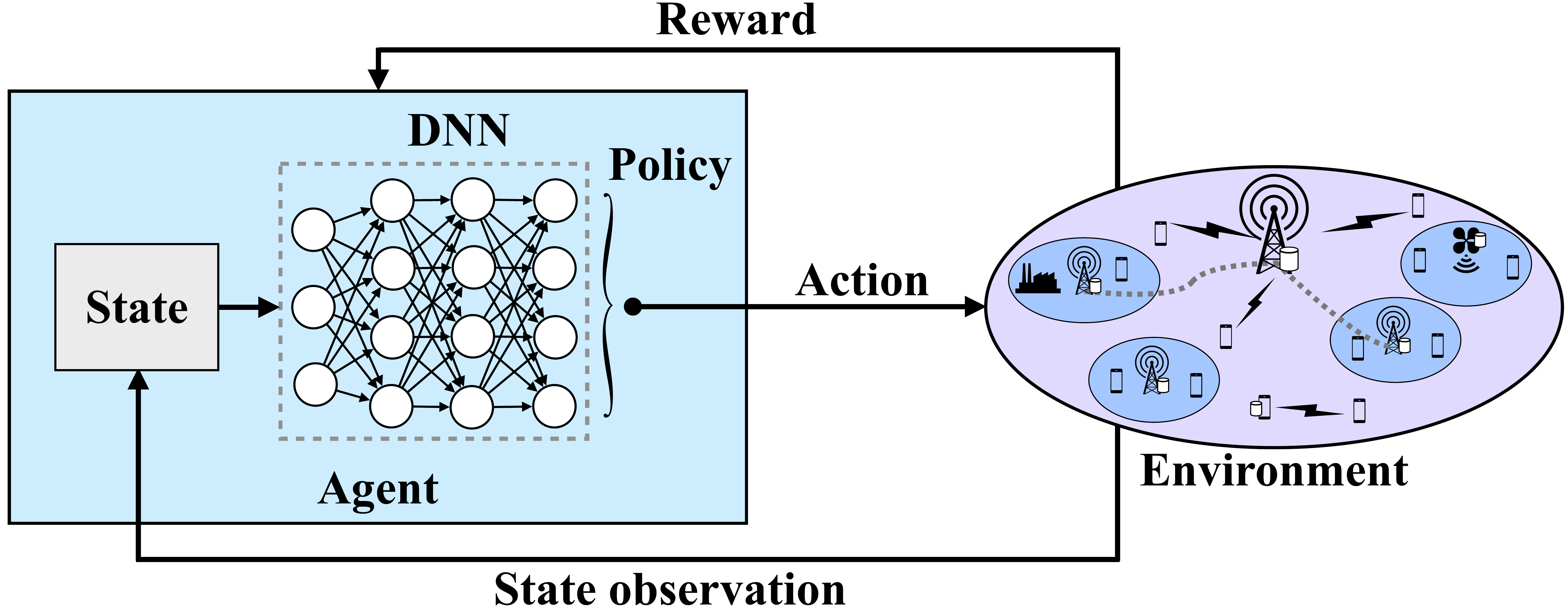}
\caption{Deep learning architecture in the context of a mobile edge environment.}
\label{DL}
\end{figure}
\item \textbf{Federated learning:} This approach decouples model training from requiring direct access to raw training data. In federated learning (FL) users exploit shared models trained from excessive amounts of data, without the need to centrally store it \cite{mcmahan2017aistats}. Here, devices take part as clients in a federation aiming at solving the learning task while being coordinated by a central server. Each client maintains a local training dataset that is not uploaded to the server and only computes and communicates an update to the current global model of the server. FL benefits applications where training can be based on already available data at each client, and guarantees high privacy and security levels, since attacks affect only individual devices, and not the cloud. 
\end{list4}

\subsection{Contributions}\label{contr}
In recent years, the integration of ML in wireless networks has offered tremendous potential towards achieving the targets of 6G networks. In addition, edge caching has been considered an enabling technique for an overall improvement of wireless network performance, through offloading and reduced number of content fetching from remote web servers. There have been several surveys focusing on either the synergy of ML and wireless networks \cite{zhang2019cst, gunduz2019jsac} or the gains of edge caching through traditional optimization approaches \cite{li2018cst,yao2019cst}. 

Meanwhile, most studies provide brief overviews on ML-aided edge caching and non-exhaustive lists of relevant works \cite{mcclellan2020as,zhu2018network}. Currently, only a few works offer a holistic view on ML-aided edge caching. 
The survey in \cite{anokye2019zte} investigates ML-based proactive caching, highlighting the improvement in small cells and UAV-aided networks. However, the majority of the reviewed works are non-learning-based. Another survey studies DL for edge caching, presenting the major DL categories and the basic caching principles \cite{wang2020informatics}. Still, the survey mostly focuses on describing DL operation in a more general manner while discussion of the various works often lacks the details on the networking environments and performance evaluation. {The use of artificial neural networks (ANNs) for wireless network optimization is examined in \cite{chen2019cst}, including the area of edge caching. However, the tutorial dedicates only one part for ANN-aided edge caching and includes a small number of relevant works.} Finally, the survey in \cite{sheraz2020comst} provides an overview of AI-based techniques for wireless caching, including supervised, unsupervised, reinforcement and transfer learning. Various challenges are highlighted, such as the dynamic environment due to mobility and fading. Still, that work does not provide a broad view of RL-based solutions, representing the main category of ML that can handle the volatility of mobile edge networks.

Considering the importance of edge caching in the context of 6G and the high interest in developing RL- and {deep reinforcement learning\footnote{{Deep reinforcement learning combines RL and neural network based function approximators in order to tackle the curse of dimensionality.}} (DRL)}-aided solutions, %\todo{[changed it from DL to DRL for keeping coherence and for being different than \cite{chen2019cst}]}
this survey provides an exhaustive list of RL-based edge caching solutions and a spherical view on their application in mobile edge networks. Thus, it is the first survey, focusing only on RL-aided edge caching, discussing the recent contributions on this field. More specifically, our contributions include:

\begin{list4}
    \item Various cache-aided wireless networks are covered, from fixed cellular topologies and cooperative architectures to fog-based approaches and vehicular networks.%, giving details on the role of cache-aided nodes and the use of RL.
    \item RL-based edge caching schemes are discussed, presenting in detail the learning algorithms, possible caveats for their implementation and the gains they bring to the network.
    \item A classification of the RL solutions is conducted, according to their performance target, i.e. energy, spectral and caching efficiency, delay and Quality-of-Experience (QoE).
    \item Open issues are highlighted, stimulating further research on this field and discussing the interplay with other aspects of wireless networks, such as physical-layer and multiple access issues, security and network volatility.
\end{list4}

\subsection{Organization}
The structure of this survey is as follows. In Section~\ref{fixed}, we present RL-aided edge caching in fixed cellular networks, while Section~\ref{fog} focuses on F-RANs. Then, Section~\ref{coop} includes cooperative approaches, as well as local caching at mobile devices. Subsequently, highly mobile and flexible network topologies are discussed in Section~\ref{veh} and Section~\ref{uavs}, %i.e., vehicular and UAV-aided networks.
Open issues in RL-based edge caching are presented in Section~\ref{open} and finally, conclusions are given in Section~\ref{conclusions}. %\todo{{[Due to the wide variety of scenarios, I would possibly consider including ToC, like this one: \href{https://people.kth.se/~kallej/papers/network_arc19yang.pdf}{https://people.kth.se/~kallej/papers/network_arc19yang.pdf}]}}
\tableofcontents

\begin{table}[H]{
\centering
\caption{List of abbreviations}
\label{abbreviations}
\begin{tabular}[t]{  m{5em} |  m{20em}  } 
  %\toprule
  %\textbf{Symbol} &  \\
  %\hline
  A3C & Asynchronous advantage actor-critic \\
  AC & Actor-critic \\
  AoI & Age of information \\
  AP & Access point \\ 
  BBU & Baseband unit \\
  BLA & Bayesian learning automata \\
  BS & Base station \\ 
  C-RAN & Cloud radio access network \\
  %CCAM & Collaborative caching and auction matching \\
  CDM & Content delivery market \\
  CNN & Convolutional neural network \\ 
  CPU & Central processing unit \\
  CRNN & Convolutional recurrent neural network \\ 
  D2D & Device-to-device \\ 
  DCA & Deterministic caching algorithm \\
  DDPG & Deep deterministic policy gradient \\
  DDQN & Double deep Q-network \\ 
  DL & Deep learning \\ 
  DNN & Deep neural network \\
  DQL & Deep Q-learning \\
  DQN & Deep Q-network \\
  DRL & Deep reinforcement learning \\
  DTS & Double time-scale \\
  %ECC & Expected correlation coefficient \\
  ELSM & Echo liquid state machine \\
  eMBB & Enhanced mobile broadband \\
  EMQRN & External memory-based recurrent Q-network \\ 
  F-RAN & Fog radio access network \\
  FIFO & First-in first-out \\ 
  FL & Federated learning \\
  HetNet & Heterogeneous network \\
  ICRP & Individual content request probability \\
  IoT & Internet-of-Things \\
  ITU & International telecommunication union \\
  KNN & K-nearest neighbors \\
  %LFM & Least frequently caching and matching \\
  LFU & Least frequently used \\
  LMP & Local most popular \\
  LRU & Least recently used \\
  LSTM & Long-short-term memory \\ 
  M2M & Machine-to-machine \\
  MAB & Multi-armed bandit \\
  MDP & Markov decision process \\
  MEC & Mobile edge computing \\
  MIMO & Multiple-input multiple-output \\
  MINLP & Mixed integer non-linear program \\
  ML & Machine learning \\ 
  mMTC & Massive machine type communications \\
  {MNO} & {Moblie network operator} \\
  MVNO & Mobile virtual network operator \\
  NOMA & Non-orthogonal multiple access \\ 
  {PLS} & {Physical-layer security} \\
  PPO & Proximal policy optimization \\
  Q-LCCA & Q-learning collaborative cache algorithm \\
  QoE & Quality-of-Experience \\
  QoS & Quality-of-Service \\
  %RAN & Radio access network \\ 
  RB & Radio bearer \\ 
  RL & Reinforcement learning \\ 
  RLNC & Random linear network coding \\
  RRH & Remote radio head \\
  RRM & Radio resource management \\
  RSU & Road Side Unit \\
  SAE & Stacked auto-encoder \\
  SDDPG & Supervised deep deterministic policy gradient\\
  SDN & Software-defined network \\
  SNR & Signal-to-noise ratio \\
  UAV & Unmanned aerial vehicle \\
  UCB & Upper-confidence bound \\
  URLLC & Ultra-reliable and ultra-low latency \\
  V2I & Vehicle-to-infrastructure \\
  V2V & Vehicle-to-vehicle \\ 
  V2X & Vehicle-to-everything \\
  VFA & Value function approximation \\
  VR & Virtual reality \\
  WMMSE & Weighted minimum mean square error \\
  %\bottomrule
\end{tabular}\label{notation}}
\end{table}

\section{Fixed Cellular Networks}\label{fixed}
Cellular architectures comprising fixed base stations (BSs) represent a major field where edge caching can alleviate the burden of excessive data traffic from users residing in their coverage area. Two types of architectures will be discussed, {namely}, single-cell topologies and multi-cell topologies where different tiers overlap. 

\subsection{Single-cell topologies}

There have been various works studying the performance of RL-aided edge caching in single-cell networks, presenting scalable solutions that can be employed to more complicated settings, after necessary modifications.

\subsubsection{Energy efficiency}
The paper in \cite{yang2019icc} studies resource allocation in cache-aided MEC networks to guarantee sufficient communication, caching and computing capabilities for intensive computational tasks with stringent latency constraints. Towards satisfying such constraints, the joint optimization of task offloading, as well as cache, computation and power allocation is formulated as a mixed integer non-linear program (MINLP) problem. Then, resource allocation is modeled as a Markov decision process (MDP) and a DRL framework is proposed, enabling the users and the access point (AP) to learn from historical data and increase the efficiency of resource allocation. Furthermore, DRL allows a quasi-optimal solution to be obtained with low-complexity, even under the large state space of the MDP. From the simulations, it was shown that DRL provides improved energy consumption, as the AP caching capability increases, while for increasing computation capacity, the energy consumption performance is near-optimal. Also, comparisons with other benchmark schemes without caching capabilities and different task computation strategies emphasize on the important energy gains of the MDP-based DRL algorithm.

Resource allocation for improved energy performance in cache-aided networks has also been investigated in \cite{yang2020icc, yang2020twc} where non-orthogonal multiple access (NOMA) has been employed. NOMA enables multiple users to simultaneously offload tasks to APs, operating as edge computing servers, achieving latency reduction. At the same time, the caching of computational results reduces the stress under increased computation requests, as these results can be requested at a different time by other users enjoying the same application. So, in this context, the problem of jointly optimizing task offloading, computation resource allocation and caching decisions is tackled by employing a long-short-term memory (LSTM) network. LSTM improves the exploration-exploitation trade-off when predicting task popularity. For the resource allocation procedure, a single-agent Q-learning algorithm is utilized while Bayesian learning automata (BLA) multi-agent Q-learning handles task offloading. Extensive simulations are conducted, depicting the high prediction accuracy of LSTM. Comparisons of the single-agent Q-learning algorithm with three benchmark schemes, i.e., local computation at the mobile users, computation only at the AP and computation in a non-cache-aided networks highlight important energy savings for the proposed RL algorithms, as caching and computation capacities are increasing. Finally, it is observed that BLA multi-agent Q-learning offers improved energy consumption performance over an algorithm that does not employ BLA for task offloading.

Focusing on optimizing the content update strategy in small BSs with limited cache capacities, the works in \cite{sadeghi2018icassp,sadegh2019jsac} consider a more general framework characterized by random resource availability and content requests. More specifically, time-varying and stochastic costs are assumed, being associated with file fetching from the cloud, incurring scheduling, routing and transmission costs. Also, cost includes memory and energy consumption due to caching at the small BS. Two different cases are examined where in the first case, costs and content popularity follow known and stationary distributions and a dynamic programming problem is formulated and solved through value-iteration-based RL. The second and more practical case considers limited cache capacity and unknown cost distributions and an online low-complexity Q-learning solver is employed to determine the optimal content update strategy. The caching versus fetching trade-off is evaluated for both cases with varying mean values for the cost of caching and fetching. It is revealed that the online Q-learning without a priori knowledge of the statistical properties of the costs and content popularity offers almost the same average cost performance with value-iteration-based RL, while its efficiency in cache-fetch decision-making is emphasized in both stationary and non-stationary environments. 

\subsubsection{Caching efficiency}
Targeting to improve the data offloading and cache hit rate performance, the paper in \cite{zhong2020tccn} presented DRL with deep deterministic policy gradient (DDPG)-based training \cite{lillicrap2019arxiv} and the Wolpertinger policy \cite{dulac2015arxiv}, relying on three entities. First, an actor function, receiving as input the cache state and the content requests and providing a proto-actor from the set of valid actions. Then, K-nearest neighbors (KNN) mapping is employed, expanding the proto-actor to a set of valid actions from the action space. Finally, a critic function refining the actor in order to select the action with the highest $Q$-value from the expanded KNN set. Performance evaluation for centralized caching and comparisons with least recently used (LRU), least frequently used (LFU), and first-in first-out (FIFO) policies reveal that actor-critic (AC) DRL can improve the cache hit rate in the short-term and avoid cache hit rate variations in the long-term. 

The issue of accurately replacing the cached content at a BS that does not know the content popularity is investigated in \cite{wu2019cl}. This problem is cast as an MDP where the BS cache status and the user requests represent the state space, while the decision of either keeping the current files or replacing them with updated versions define the action space. So, the authors propose an LSTM and external memory-based recurrent Q-network (EMQRN) algorithm to enhance the cache hit rate. The proposed algorithm is evaluated and compared with LRU and FIFO without learning capabilities and the deep Q-network (DQN) algorithm of \cite{mnihnature2015}. From the results, it can be seen that EMQRN leads to higher reward and faster convergence, compared to DQN. Still, the implementation of EMQRN in more complicated topologies where multiple BSs cooperate and share their cache status for optimized content update strategies remains an open problem.

\subsubsection{QoE improvement}
In a network where a single caching server facilitates the transmission of short video content, a gradient-based DRL algorithm was developed in \cite{wu2019access}. Two issues are concurrently tackled, i.e., the video quality selection and the radio bearer (RB) control transmission performance. This joint problem is modeled as an MDP for the sequence of user content requests, triggering necessary RB control actions, such as setup, reconfiguration and release. For each action, the decision is made by considering the current state while training in scenarios with different parameters enables the proposed DRL to be employed in different scenarios with the same state spaces. The performance of the gradient method-based DRL is evaluated against a greedy policy serving the content request with the highest waiting time, and transmitting it at the highest video quality, and a minimum quality policy, aiming at serving the maximum number of requests by setting the video quality at the lowest level. By considering the use of additional RBs as cost and the increased level of video quality as reward, different arrival rate cases are tested and it is observed that DRL outperforms the greedy policy for rates between 1.8 to 2.4 while for rates below 1.6 the greedy method provides better performance but at the cost of higher complexity. An open issue remaining is the generalization of the proposed gradient-based DRL for more varied communication scenarios.

Focusing on content-centric caching for improved QoE, the authors in \cite{he2019cim} developed a DRL-based decision making model, employing DNN for Q-value estimation. Optimization considered both latency and storage costs, outlining the negatively proportional relationship of the two metrics to QoE. As there exists a conflict between these two performance metrics and the network operates in a dynamic environment, DNN might not effectively estimate the Q-value. Thus, DRL relied on fixed target network (F), experience replay buffer (E), and adaptive learning rate (L) for improved stability, leading to FEL-DRL.  More specifically, fixed target network leads to stable convergence by using another neural network with fixed parameters which are periodically according to the values of the estimated network. Meanwhile, experience replay avoids temporal correlation among different training episodes, creating a dataset from the agent's experience and randomly using data batches for network training. Comparisons using Matlab$^\copyright$ and TensorFlow showed that FEL-DRL achieves an average QoE score of 64, while DRL provided a score of 62. Finally, the rest of the algorithms, i.e. AC-DRL, FE-DRL and RL provided QoE values below 60 which is considered as the threshold for user satisfaction.

\subsection{Multi-cell topologies}

More complicated multi-cell networks have been investigated in various works, highlighting the gains of edge caching at different nodes. 

\subsubsection{Age of information reduction}
The authors in \cite{ma2020icc,ma2020twc} focus on the update scheduling for minimizing the age of information (AoI) of the cached content, in a two-tier heterogeneous network (HetNet) where multiple small cells act as content servers, delivering dynamic content to users. AoI quantifies how much time has passed from the moment that the current file version has been generated \cite{yates2017tit, yates2021cst}. The content caching problem is formulated as a constrained MDP and enforced decomposition is employed to update the dynamic contents in the caches. AoI is minimized by using multiple queues to monitor user request, assuming that each request arrives prior to the desired file download time. As the state space of the different constrained MDP subproblems can be large, DRL agents are trained to derive the optimal content update policy. Performance evaluation results, using PyTorch suggest that DDPG-based DRL offers improved convergence and reduced AoI. More specifically, for the single dynamic content scenario, DDPG provides improved convergence, compared to DQN \cite{mnihnature2015}, while for multiple dynamic contents, the average AoI is reduced by 30\% versus periodic update without considering the user request queues \cite{yatesisit2017}.

\subsubsection{Delay reduction}
The reduction of transmission delay and cache replacement cost in the long-term, in two-tier small cell networks is studied in \cite{zhang2019iccc}. More specifically, wireless coded caching is employed to distribute coded fractions of a file at different network nodes, thus addressing the need for large caches. As a first step to provide improved caching, historical user file requests are exploited for predicting the future ones. In the next phase, a supervised deep deterministic policy gradient (SDDPG) approach based on both supervised learning and DRL is employed to solve the wireless coded caching problem. Aiming at accelerating the learning process, supervised learning is invoked to pre-train the neural network by considering the solution to an approximate cost minimization problem at each slot. Performance evaluation results, highlight the the capabilities of SDDPG in reducing the total network cost, compared to a short-term optimization cost scheme while it exhibits a small performance gap compared to the performance upper-bound of knowing the actual number of requests.

The MAB framework is adopted in \cite{xu2020twc}, mapping the reward to transmission delay reduction, compared to the case without caching. Assuming unknown user preferences, the proposed collaborative caching schemes aim at minimizing the accumulated transmission delay over a finite time horizon. This paper extends the work in \cite{xu2018gc} which presented distributed and collaborative multi-agent MAB algorithms in stationary environments. Here, two stationarity cases are investigated for the file library and user preferences. For the stationary case, a fixed file set and time-invariant user preferences are considered and two high-complexity MAB algorithms are presented. Their regret performance is bounded by $\mathcal{O}(\log T_{\textrm{total}})$, where $T_{\textrm{total}}$ denotes the total number of time-slots. Meanwhile, a lower complexity and distributed MAB solution is developed, considering that each small BS acts independently. Judging from the performance of these algorithms, an edge-based collaborative multi-agent MAB algorithm is proposed, relying on coordination graph edge-based reward assignment. Then, in the non-stationary case, the file set and user preferences dynamically vary and thus, modified multi-agent MAB algorithms are given. More specifically, the exploration duration is reduced by assigning larger initial values to the joint actions of adding new content to the varying file set in each time-slot. Also, the upper confidence bound (UCB) terms are modified to accommodate that the small BSs are unaware of the reward upper bound. Simulations for both cases show that the proposed MAB algorithms significantly reduce the delay, compared to policies, such as LRU and LFU, while narrowing the performance gap compared to an oracle greedy algorithm for a wide range of communication distance, cache size and user mobility parameters.

The delay minimization is also the focus in \cite{wei2018icc} where AC-based DRL joint user scheduling and content caching is proposed. In greater detail, the actor adopts stochastic caching, abiding to the Gibbs distribution and parameters are updated through gradient ascent by observing the environment states. The critic evaluates the actor policy and the resulting awards, in terms of delay. For this purpose, a DNN is used for value function approximation and gradient estimation. The convergence of the AC-based DRL scheme is evaluated in a two-tier network for different learning rates for the actor and the critic, highlighting that a low actor learning rate leads to improved convergence. Also, comparisons are presented with AC-based DRL without caching and AC-based DRL without scheduling. Results indicate 40\% and 56\% higher total rewards through the proposed joint user scheduling and caching scheme over the standalone scheduling and caching schemes, respectively.

The joint optimization of delay and blockchain-based security has been presented in \cite{li2020iotj}, focusing on (machine-to-machine) M2M communication. Since blockchain systems require increased time to complete the smart contracts,  delay requirements might not be met \cite{liu2019tii}. So, in this study, system performance is enhanced by developing a dueling DQN for optimal decision-making, regarding caching, computing and security. The dueling architecture allows the DDQN to efficiently learn the action value through the separate estimation of the state value and the reward of each action, including higher caching reward, reduced data computation overheads, and efficient blockchain processing. Performance comparisons against conventional DQN, greedy-based strategy and random selection for different cache and block sizes, delay constraints and a varying number of machine-type devices, showed reduced latency, and higher rewards when dueling DQN is employed.

\subsubsection{Caching efficiency}
An AC-based DRL policy is also employed in a network with multiple BSs, towards providing decentralized edge caching \cite{zhong2020tccn}. In this setting, the main goal of the learning-based approach is cache hit rate performance improvement. Comparisons with non-learning-based LRU, LFU and FIFO policies shows that AC-based DRL efficiently finds the popular contents and reduces the transmission delay by considering user location and enabling the inter-BS communication in order to avoid caching of the same content when their coverage areas overlap. 

The optimization of content caching and delivery policy under non-stationary content libraries is the subject of \cite{zhang2020tcom}. Aiming to maximize the network utility consisting of backhaul traffic offloading, cache hit rate, as well as the content retrieving and delivery metrics, a user-assisted RL algorithm is given. This algorithm exploits the caches of the users in order to relieve the small cells during peak hours. So, the BSs' caches are divided in two parts where in the first part, new content from the users' caches is stored, while using the second part for content server updates. The content caching and delivery is formulated as a MAB problem, accommodating the spatio-temporal dynamics of user requests. In this context, the content library is modeled as a system with multiple arms with unknown and stationary rewards. A central unit sequentially determines content caching, based on the trade-off of exploring possibly popular files that are rarely cached and exploiting the empirical knowledge of caching content, yielding the highest rewards up to this round. The proposed content caching and delivery algorithm operates in three phases, where in the first phase, content delivery takes place, then, in the second phase, one part of the small cell caches is updated with content from the users and finally, in the third phase, the other part of the BSs' caches is updated from the content server. Performance evaluation includes comparisons with benchmark schemes operating without a user-assisted phase. It is observed that the MAB-based user-assisted algorithm is more robust against the spatio-temporal variations of content requests and benefits from the exploitation of the users' caches. Nonetheless, it is stated that providing regret performance guarantees is challenging, since this works assumes caching of multiple contents at different small cells, coordinating with each other during the last phase and possibly replacing content from the users' caches from the second phase.

Distributed content placement was studied in \cite{sengupta2014iswcs} in a dense small cell network, aiming to alleviate the traffic load from the backhaul infrastructure. The authors showed that the problem of optimal content placement is NP-hard, independently of whether or not the small BSs know the file popularity profiles. Thus, a learning-based coded caching solution is proposed where the small BSs are employed to learn the file popularity profiles, using the content request historical data. The learning framework takes into account the connectivity of users with the small BSs and relies on combinatorial MAB. More importantly, the MAB-based learning framework is able to adapt to the dynamicity of content popularity over time, based on the trade-off among the exploration of caching new files and find their popularity versus the exploitation of caching files with already-known high popularity. Regarding the reception of distributed cached contents, rateless coding is adopted, guaranteeing the decoding of the original file, as long as a specific fraction of the coded symbols is received. Performance evaluation depicts the advantage of MAB-based distributed caching in yielding higher rewards, compared to a local caching scheme  neglecting the network connectivity status and an uncoded caching scheme.

In \cite{thar2019access} the authors aim at improved edge caching in networks where infrastructure providers lease their physical resources, in the form of BS storage and backhaul capacity to mobile virtual network operators (MVNOs). By investigating the joint optimization of cache leasing and content popularity prediction from the MVNOs' perspective towards maximizing their profits, a Q-learning algorithm is presented to provide DL models with optimized hyper-parameters. The generated DL models are employed to predict the parameters of content popularity, namely future cache demand and request count. Using this information, the DL models compile lists with content that should be cached at the BS. Performance evaluation focuses on the cache hit probability and backhaul usage and three different configurations for the unknown layer of the DL model, i.e., convolutional neural network (CNN), LSTM and convolutional recurrent neural network (CRNN). The results on feature selection suggest that LSTM provides superior training and validation accuracy performance with the least amount of training time. Finally, after evaluating different configurations, the best LSTM models are compared versus the optimal and randomized caching schemes, showing that on average, the best LSTM model offers a 16\% cache hit probability improvement, compared to 12\% by the randomized scheme, while it reduces backhaul usage by 17\%, compared to 12\% by the randomized scheme.

RL-based meta-learning with enhanced searching space design and autonomous DL model generation, as presented in \cite{thar2019access} with optimized hyper-parameters are investigated in \cite{thar2019cnsm}. The proposed solution comprises two parts. In the first part, a cloud-based master meta-learner provides the DL models and decides on the best-suited one. The second part involves a slave meta-learner located at each small BS, using the best DL model for popularity prediction after tuning its parameter through localized information. Simultaneously, the slave meta-learner provides feedback to the master meta-learner on the prediction accuracy, triggering the latter to explore a different model, in case of suboptimal results. Performance evaluation is conducted by implementing the RL-based meta-learning scheme, using Tensorflow \cite{tensorflow} and Keras \cite{keras}. Comparisons show 10\% and 30\% cache hit rate improvement over the scheme in \cite{thar2019access} and randomized caching, respectively.

Another work aiming to increase the net profits of mobile network operators (MNOs) through a DRL framework is presented in \cite{liu2019access}. Here, the content is not cached at the BSs but on the contrary, it is proactively pushed and cached at the users' devices. In order to achieve this, content pushing and recommendation for the users are investigated, resorting to RL for predicting the individual user behavior. Nonetheless, the joint problem of proactive pushing and recommendation is characterized by large action and state spaces and thus, a decomposition approach is followed. In greater detail, the recommendation subproblem focuses on increasing requests and providing revenue opportunities. Meanwhile, the pushing subproblem targets aims at minimizing the transmission delay. Considering the inter-dependency among the two subproblems, a double deep Q-network (DDQN) relying on the dueling architecture of \cite{wangicml2016} is employed. Extensive simulations are conducted to assess the performance of the dueling DDQN versus other learning algorithms, including DDPG \cite{lillicrap2019arxiv}, advantage AC \cite{mnih2016icml} and proximal policy optimization (PPO) \cite{schulman2017arxiv}. Results highlight that dueling DDQN converges much faster while solving the recommendation subproblem and provides the highest rewards. More specifically, even though, PPO provides almost the same reward as dueling DDQN, it requires around 43\% additional training sessions, compared to the dueling DDQN. Meanwhile, the pushing policy is able to exploit the mobility pattern of the users and the propagation characteristics, proactively pushing content under favorable channel conditions.

\begin{figure}[t]
\centering
\includegraphics[width=\columnwidth]{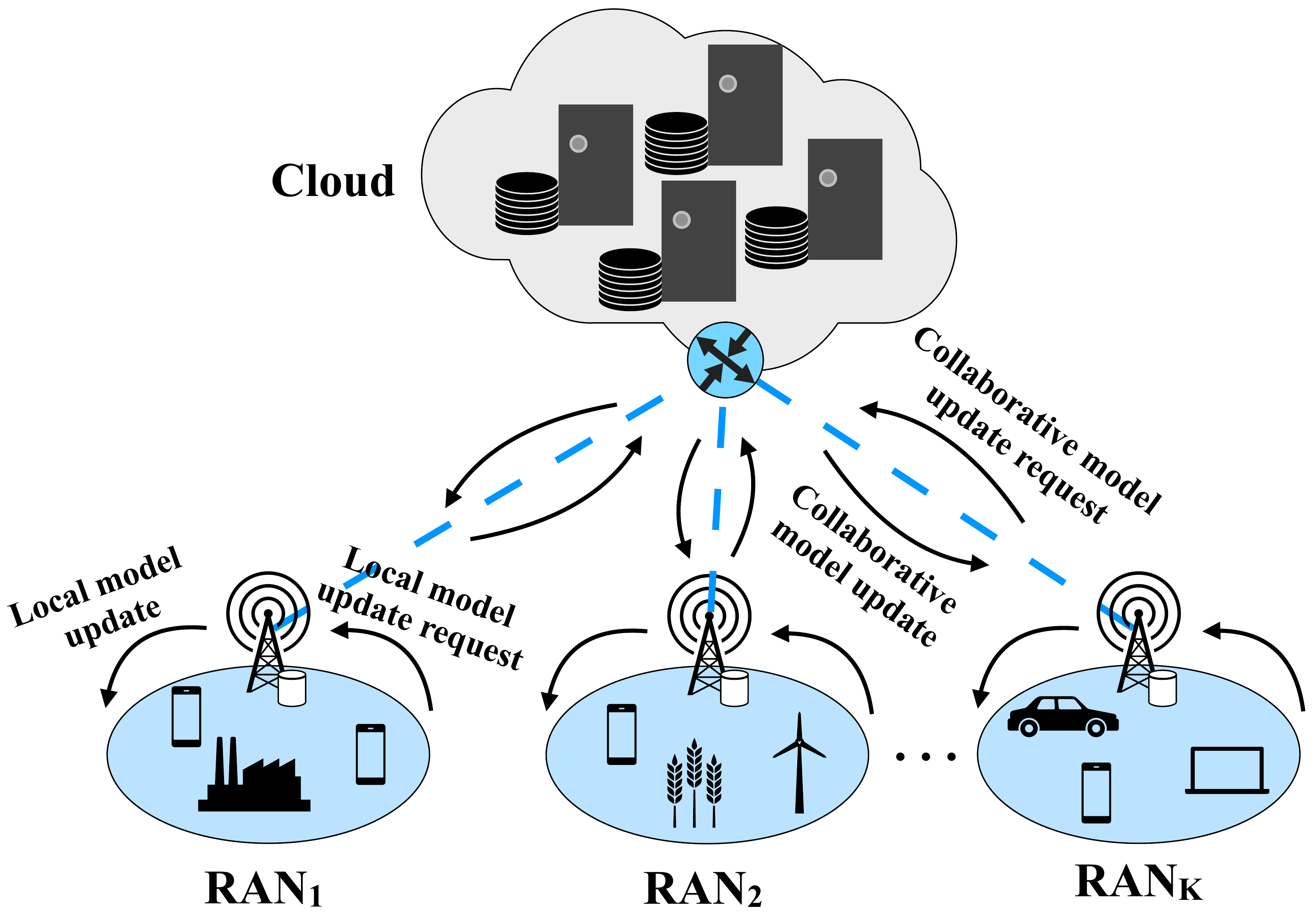}
\caption{Federated learning architecture for a multi-cell network, comprising $K$ mobile edge environments.}
\label{FL}
\end{figure}

In \cite{wang2019network}, DRL-based network resource management for achieving better cache hit rate and computation offloading is presented. Nonetheless, in mobile edge networks, the amount of data, parameters and performance targets calls for distributed DRL agent training. Two important aspects are discussed, determining the distributed DRL architecture that should be adopted. First, maintaining a DRL agent in every network node can provide improved performance, but in practice, training will struggle due to differences in task load and network states, as well as time constraints and data unavailability. Second, the distributed DRL architecture should be able to overcome data imbalance and alleviate privacy concerns. So, FL is employed for distributed DRL agent training, reducing communication costs and offering improved privacy and security \cite{mcmahan2017aistats}. Fig.~\ref{FL} shows a multi-cell topology with $K$ RANs adopting FL to optimize their operation by exploiting local model updates to improve the efficiency of the global collaborative model which is available to all the members of the federation. The proposed In-Edge AI, reduces the need to constantly uploading data in the uplink as FL relies on locally stored data and only calculate updates to the global model of the coordinating central node. Simulations compare the caching performance of DDQN with FL and centralized DDQN without FL, as well as LRU, LFU and FIFO. It is observed that DDQN with FL provides almost the same hit rate performance as centralizes DDQN and outperforms LRU, LFU, and FIFO. In addition, even though the simulated wireless topology is assumed to support the upload of the large amount of training data in the centralized DDQN approach, in practical networks, delay will severely degrade its performance while the small volume of data for FL-based training through the global model updates will be slightly affected.

In cases where user preferences and mobility patterns are unknown, the authors in \cite{guo2019access} proposed a temporal-spatial recommendation policy addressing the issue of non-peaky local content popularity. This policy leads cellular users to request their desired files in an efficient way, i.e., during specific time-slots and through appropriate BSs. Since a limited number of user requests might hinder local popularity prediction, a Bernoulli mixture model is adopted to learn user preference and request probability. Then, the recommendation and caching policies are jointly optimized by harnessing RL algorithms. Nonetheless, this joint problem is characterized by large state and actions spaces. Thus, it is decomposed into three subproblems, related to user preference and file request probability estimation with or without recommendation, caching policy optimization, independently of recommendation, and finally, recommendation policy optimization by relying on DRL. Performance evaluation for a network with three BSs and a library of 200 files and comparisons with four baselines schemes are presented. Baselines schemes included, random recommendation, no recommendation, global recommendation, based on the estimated global file popularity and local recommendation where file recommendation in each cell is based on the aggregating the estimated local file popularity. It is shown that DRL overcomes possible user and aggregated preference estimation errors, better adapting the caching policy to user mobility through improved recommendation, compared to the other schemes.

\subsubsection{Spectral efficiency}
Pushing and caching popular services in a three-tier network consisting of broadcast BSs, cellular BSs and routers is studied in \cite{fang2019vtc}. The broadcast BS is responsible for delivering the services and the caches of the routers are used to bring the content closer to the users. In case, a user does not receive its desired content, the router might act as a relay to establish connectivity with another router or a handover to a cellular BS takes place. Targeting the maximization of the equivalent network throughput, service scheduling is modeled as an MDP. Since the large state space entails excessive complexity, deep Q-learning (DQL) is used to find the optimal policy for each state and maximize the cumulative award. Towards addressing the large state space issue, the Q-function is approximated and modified using experience replay and target value update in several time steps, over updating in each time step. Performance comparisons with the algorithm of \cite{zhang2018bmsb} and centralized caching with dynamic programming shows that for different Zipf factors, characterizing the content popularity, DQL provides significantly higher equivalent throughput, especially for Zipf values below 1.5.

The issue of enhanced edge caching performance through accurate content recommendation, based on personalized preferences is addressed in \cite{cheng2019tcom}. In greater detail, localized caching is presented, relying on the individual content request probability (ICRP) for content placement optimization and throughput maximization. This scheme is based on Bayesian learning, namely constrained Bayesian probabilistic matrix factorization, considering rating matrix imbalances towards improving the prediction accuracy of unknown content ratings. The output of this process facilitates the evaluation of personal preferences to obtain the ICRP. In the next step, RL exploits the ICRP and physical distance among caches and users for content placement, resulting in a deterministic caching algorithm (DCA). Further extensions to DCA are provided by presenting device-to-device (D2D) cooperation for reduced transmission delay and improved ICRP estimation. DCA is compared against random caching and probabilistic caching in terms of root mean square error (RMSE) prediction performance, hit rate and system throughput. From the comparisons, it is observed that DCA offers 90\% increased throughput compared to random caching, while D2D cooperation reduces the delay by 15.5\% over the non-D2D case.

The work in \cite{garg2019icassp} studied the issue of content placement at BSs towards maximizing the average success probability of the transmission. The authors considered a network with BSs located as a two-dimensional homogeneous Poisson point process (PPP) while the content popularity was assumed to be known and Zipf distributed. Then, by formulating the cost function using the average success probability, an online Q-learning approach was presented and evaluated for both small and large action spaces, depending on the number of cache size, content size and popularity profile set cardinality. Results indicate that for small action spaces, Q-learning converges to the optimal policy after 13 iterations while significantly more iterations are needed for convergence for large action spaces, i.e., around $2\times10^3$ iterations.

Aiming to alleviate the impact of tidal effects in mobile networks, i.e., increased network load in peak hours and low bandwidth utilization in idle periods, the authors in \cite{qian2020jsac} propose joint pushing and caching to proactively transmit user content when the network is underutilized. This joint problem involves a transmission cost function, representing bandwidth fluctuation. The minimization of bandwidth fluctuation leads to improved bandwidth utilization and energy efficiency while avoiding duplicate data transmissions. So, the authors propose to exploit hierarchical RL for tackling joint pushing and caching and resort to decomposing the problem into two subproblems. The first subproblem is related to user cache optimization, employing Q-learning value function approximation to mitigate the impact of large state and action spaces. For the second subproblem, DRL is used to improve the performance of BS caching and tackle dimensionality issues. In order to evaluate the performance of the proposed hierarchical RL, comparisons with other policy-based schemes, such as LRU, LFU and local most popular (LMP) in three scenarios, i.e., caching at the BS, caching at the users, joint BS and user caching, are presented. It is observed that hierarchical RL outperforms the other policies in all three cases, while its advantage is significantly increased in the joint BS and user caching, since both the wired and wireless network parts are efficiently utilized.

The joint allocation of networking, caching, and computing resources for supporting smart cities applications is examined in \cite{he2017cm}. Assuming a dynamic virtualized networking environment where MVNOs manage multiple BSs, MEC servers and content caches, an excessive number of system states exists and traditional optimization faces difficulties in derviving the optimal policy. As a result, DRL is invoked, using DQN for Q-value function approximation, determining the resource allocation of networking, caching and computing resources. The MVNO's revenue, formulated as a function of the access link signal-to-noise ratio (SNR), the MEC server computation capability, and the cache state represents the system reward. Simulations are performed, using TensorFlow and DRL is compared to alternative versions without caching, MEC offloading or virtualization. Results suggest that the proposed DRL offers a significantly higher total utility, independently of learning rate and the number of required central processing unit (CPU) cycles per task. 

\subsubsection{QoE improvement}
The resource allocation and user association problems in a network providing live video streaming service is the subject of \cite{chou2020icc}. As the maximization of the QoE is prioritized in such networks, DDPG-based learning algorithm is presented, as an alternative to traditional Lagrangian-based optimization. Initially, an optimization problem is formulated and found to be non-linear and NP-hard. In order to convert it to a linear problem, they focus on the binary decision variables, corresponding to caching content at a BS and receiving a user request for a specific video quality from a BS. More specifically, the DDPG-based algorithm alternatively keeps one variable fixed and then relaxes both binary variables to be continuous to find a near-optimal solution. However, it is observed that in the user association/video quality subproblem, the sub-gradient method is inefficient and in some cases, only a locally optimal value might be obtained. So, the DDPG-based approach is employed where in the first step, it observes the state of resource utilization in the network and determines prices for each possible action. In the next step, the users are prompted to associate with the BSs, and request a specific video quality. In this way, the resource utilization is re-calculated based on the users' decisions and the QoE level is acquired as reward, facilitating the DDPG agent to evaluate the action and accordingly set the NN weights. The DDPG-based learning is evaluated against sub-gradient-based pricing \cite{Chen2020icnc} and the approach followed in \cite{dai2016icassp}. It is concluded that the proposed DDPG approach outperforms the two baselines algorithms, as it maintains higher QoE when the number of users increases, while for reduced resource availability, the baseline schemes perform more conservative allocation and provide reduced QoE.

The provision of enhanced QoE is also studied in \cite{luo2020twc}, while avoiding excessive energy consumption. A software-defined network (SDN) is investigated and a mechanism monitoring and processing several parameters related to BS cache, as well as user buffer status and video transmissions parameters, among others. The problem of optimizing the QoE and energy performance is modeled as a constrained MDP that is transformed into an unconstrained MDP by adopting the $T$-period drift-plus-penalty concept, and exploiting the fact that the optimization target should be given as a time average. The unconstrained MDP problem is tackled through the asynchronous advantage actor-critic (A3C) algorithm, employing its agents to run on a multi-core CPU with each thread processing one agent and providing a replica of the environment. Then, the globally shared parameter vector is asynchronously updated by using the cumulative gradients of multiple agents after a specific time period. A performance evaluation environment is developed using PyTorch and comparisons with DQN and traditional convex optimization are presented. It is shown that A3C exhibits faster convergence than DQN and requires half the energy for the desired QoE. Moreover, when compared for varying numbers of BSs and users, A3C always outperforms DQN while convex optimization fails to keep up with the varying network dynamics and falls behind both learning algorithms.

An additional work, focusing on QoE improvement is presented in \cite{wang2020infocom}, reducing the latency for users requesting video content and the overall backhaul usage. In order to improve these metrics, a multi-agent DRL-based caching framework is developed, treating each network edge, as a cooperative learning agent and avoiding the large action spaces of centralized single-agent approaches. The proposed multi-agent collaborative caching (MaCoCache) enables each agent to consider not only its caching strategy but also, that of its neighbors, while relying on the AC algorithm. Also, to better adapt to network dynamics and exploit historical data, LSTM is integrated with MaCoCache. The proposed caching framework is evaluated through simulations and comparisons with policy-based caching (LRU, LFU) and other learning-based alternatives, such as DRL without cooperation between agents and joint-action learners (JAL), utilizing stateless Q-learning-based caching \cite{jiang2019twc}. It is revealed that MaCoCache offers 73\%, 50\%, 21\% and 14\% latency and 103\%, 98\%, 59\%, 26\% backhaul cost reduction versus LFU, LRU, DRL and JAL, respectively, as well as 13\% and 7\% improved edge hit ratio, compared to DRL and JAL, respectively.

Table~\ref{fixed_table} includes the studies on RL-aided caching in fixed cellular topologies, highlighting their main performance targets and the adopted RL approach.

\begin{table}[!htbp]
\caption{List of works focusing on reinforcement learning-aided caching for fixed cellular networks.}
\label{fixed_table}
\centering
{\begin{tabular}[t]{m{7em}<{\raggedright} m{4.4em}<{\raggedright} m{6.9em}<{\raggedright} m{6.7em}<{\raggedright}}
\toprule
\textbf{Reference}	& \textbf{Network topology} & \textbf{Performance target}  & \textbf{RL solution}\\
\midrule%

\vspace{0.12cm} Yang et al.~\cite{yang2019icc} &\vspace{0.12cm} Single-cell &\vspace{0.12cm} Energy consumption &\vspace{0.12cm} DRL \\%

\vspace{0.12cm}Yang et al.~\cite{yang2020icc, yang2020twc} &\vspace{0.12cm} Single-cell NOMA &\vspace{0.12cm} Energy consumption &\vspace{0.12cm} BLA Q-learning \\%

\vspace{0.12cm}Sadeghi et al.~\cite{sadeghi2018icassp, sadegh2019jsac} &\vspace{0.12cm} Single-cell &\vspace{0.12cm} Energy, backhaul and storage costs &\vspace{0.12cm} Value-iteration-based and Q-learning\\%

\vspace{0.12cm}Zhong et al.~\cite{zhong2020tccn} &\vspace{0.12cm} Single- and multi-cell &\vspace{0.12cm} Cache hit rate &\vspace{0.12cm} AC and KNN-based DRL\\%

\vspace{0.12cm}Wu et al.~\cite{wu2019cl} &\vspace{0.12cm} Single-cell &\vspace{0.12cm} Cache hit rate &\vspace{0.12cm} LSTM and external memory-based DRL\\%

\vspace{0.12cm}Wu et al.~\cite{wu2019access} &\vspace{0.12cm} Single-cell &\vspace{0.12cm} QoE, spectral efficiency &\vspace{0.12cm} Gradient-based DRL\\%

\vspace{0.12cm}He et al.~\cite{he2019cim} &\vspace{0.12cm} Single-cell &\vspace{0.12cm} QoE &\vspace{0.12cm} DRL\\%

\vspace{0.12cm}Ma et al.~\cite{ma2020icc, ma2020twc}  &\vspace{0.12cm}
Multi-cell &\vspace{0.12cm}  AoI &\vspace{0.12cm} DDPG-based DRL\\%

\vspace{0.12cm}Zhang et al.~\cite{zhang2019iccc}   &\vspace{0.12cm} Multi-cell &\vspace{0.12cm}  Delay, cache replacement cost &\vspace{0.12cm} SDDPG RL\\%

\vspace{0.12cm}Xu et al.~\cite{xu2020twc}   &\vspace{0.12cm} Multi-cell &\vspace{0.12cm}  Delay &\vspace{0.12cm} Multi-agent MAB\\%

\vspace{0.12cm}Wei et al.~\cite{wei2018icc}   &\vspace{0.12cm} Multi-cell &\vspace{0.12cm}  Delay &\vspace{0.12cm} AC-based DRL\\%

\vspace{0.12cm}Li et al.~\cite{li2020iotj}   &\vspace{0.12cm} Multi-cell &\vspace{0.12cm}  Delay &\vspace{0.12cm} DDQN\\%

\vspace{0.12cm}Zhang et al.~\cite{zhang2020tcom}   &\vspace{0.12cm} Multi-cell &\vspace{0.12cm} Cache hit rate, backhaul usage  &\vspace{0.12cm} MAB-based RL\\%

\vspace{0.12cm}Sengupta et al.~\cite{sengupta2014iswcs}  &\vspace{0.12cm} Multi-cell &\vspace{0.12cm} Cache hit rate, backhaul usage &\vspace{0.12cm} MAB-based RL\\%

\vspace{0.12cm}Thar et al.~\cite{thar2019access}  &\vspace{0.12cm}
Virtualized multi-cell &\vspace{0.12cm} Cache hit rate, backhaul usage &\vspace{0.12cm} Q-learning for LSTM/CNN/ CRNN selection\\%

\vspace{0.12cm}Thar et al.~\cite{thar2019cnsm}  &\vspace{0.12cm}
Multi-cell &\vspace{0.12cm} Cache hit rate, backhaul usage &\vspace{0.12cm} Q-learning for LSTM/CNN/ CRNN selection\\%

\vspace{0.12cm}Liu et al.~\cite{liu2019access}  &\vspace{0.12cm} Multi-cell &\vspace{0.12cm} Cache hit rate &\vspace{0.12cm} Dueling DDQN\\%

\vspace{0.12cm}Wang et al.~\cite{wang2019network} &\vspace{0.12cm} Multi-cell &\vspace{0.12cm} Cache hit rate &\vspace{0.12cm} DDQN-FL \\%

\vspace{0.12cm}Guo et al.~\cite{guo2019access}  &\vspace{0.12cm} Multi-cell &\vspace{0.12cm} Cache miss number &\vspace{0.12cm} DRL \\%

%Yu et al.~\cite{yu2020iotj} &\vspace{0.12cm} Multi-cell &\vspace{0.12cm} Cache hit rate &\vspace{0.12cm} 2Ts-DRL-FL \\%

\vspace{0.12cm}Fang et al.~\cite{fang2019vtc}  &\vspace{0.12cm} Multi-cell &\vspace{0.12cm} Equivalent throughput &\vspace{0.12cm} DQL\\%

\vspace{0.12cm}Cheng et al.~\cite{cheng2019tcom}  &\vspace{0.12cm} Multi-cell &\vspace{0.12cm} System throughput &\vspace{0.12cm} Bayesian learning and RL \\%

\vspace{0.12cm}Garg et al.~\cite{garg2019icassp}   &\vspace{0.12cm} Multi-cell &\vspace{0.12cm} Success rate &\vspace{0.12cm} Q-learning \\%

\vspace{0.12cm}Qian et al.~\cite{qian2020jsac}  &\vspace{0.12cm} Multi-cell &\vspace{0.12cm} Bandwidth fluctuation &\vspace{0.12cm} Hierarchical RL\\%

\vspace{0.12cm}He et al.~\cite{he2017cm}  &\vspace{0.12cm} Multi-cell SDN &\vspace{0.12cm} Spectral efficiency, backhaul usage  &\vspace{0.12cm} DRL\\%

\vspace{0.12cm}Chou et al.~\cite{chou2020icc}  &\vspace{0.12cm} Multi-cell &\vspace{0.12cm} QoE &\vspace{0.12cm} DDPG-based DRL\\%

\vspace{0.12cm}Luo et al.~\cite{luo2020twc}  &\vspace{0.12cm} Multi-cell SDN &\vspace{0.12cm} QoE, energy consumption  &\vspace{0.12cm} A3C-based DRL\\%

\vspace{0.12cm}Wang et al.~\cite{wang2020infocom} &\vspace{0.12cm} Multi-cell &\vspace{0.12cm} QoE, cache hit rate, backhaul usage &\vspace{0.12cm} AC and LSTM-based DRL \\
\bottomrule
\end{tabular}}
\end{table}

\section{Fog Radio-Access Networks}\label{fog}
\begin{figure}[t]
\centering
\includegraphics[width=\columnwidth]{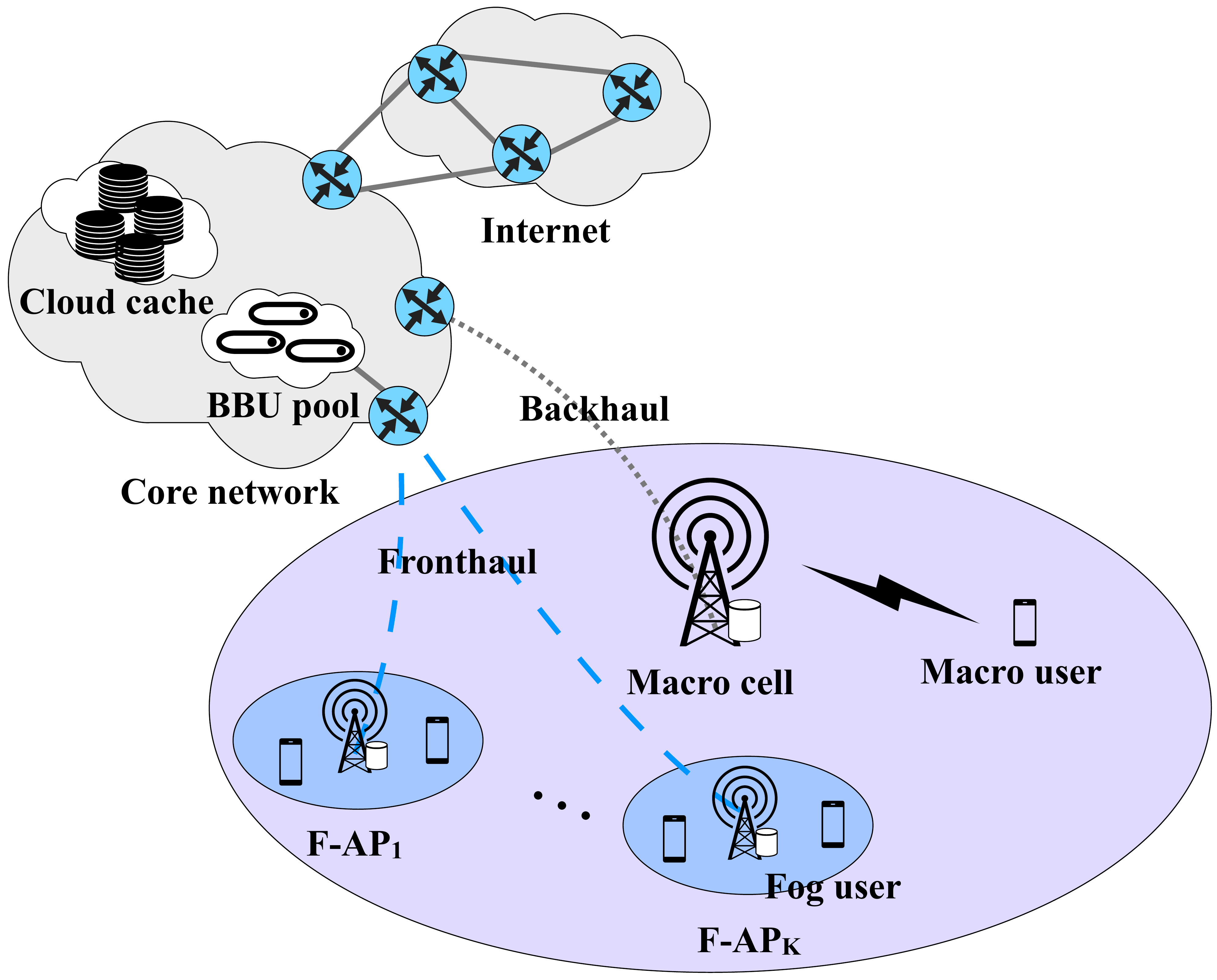}
\caption{A mobile wireless network relying on the F-RAN architecture.}
\label{fran}
\end{figure}

C-RANs provide increased flexibility to network deployment by relying on cloud-based centralized baseband units (BBUs) to perform signal processing and low-complexity remote radio heads (RRHs) for wireless access. At the same time, C-RANs may stress the fronthaul due to the massive number of content requests. Thus, F-RANs have been proposed as a promising technology towards reducing the load of the fronthaul in cellular networks. The APs in F-RANs perform a part of F-RAN architecture is illustrated in Fig.~\ref{fran} where F-APs are partly responsible for the baseband processing, while offering storage and computing resources at the edge. The adoption of RL for improving the caching and resource management efficiency in F-RANs represents an important research area that has attracted various contributions recently.

\subsection{Delay reduction}
In \cite{guo2020access}, a network comprising multiple cache-aided F-APs is considered. The goal of this study is to minimize the average delay without neglecting temporal channel variations, user mobility and varying user preferences. For this purpose, MDP is invoked to model the cache update content at the F-APs while dueling DQN is employed to solve the MDP problem without knowing the state transition probabilities. Dueling DQN estimates the state value and the rewards of the different actions in order to learn the state value, facilitating the replacement of the cached content with appropriate contents in each transmission period. Comparisons with policy-based caching including FIFO, LRU and LFU, shows that the dueling DQN delay-aware policy increases the average cache hit rate and reduces the average transmission delay for varying storage size and number of users. Meanwhile, it is noted that important performance gains can be harvested when a joint radio resource management (RRM) and cache update policy is developed, as this work assumes equal bandwidth allocation for all the users in the network.

Another work targeting latency minimization if F-RANs was presented in \cite{rahman2020tvt}. In greater detail, the authors study the joint optimization of proactive caching and power allocation in a downlink F-RAN with multiple APs and a DRL controller located at the centralized cloud. It is shown that optimizing the latency while considering the QoS of each user, the limited storage and transmit power resources of the F-APs results in a non-convex mixed-integer nonlinear fractional programming (MINLFP) problem. So, latency minimization is modeled as an MDP without a priori knowledge of the state transition probabilities and a DQN-based algorithm is developed to derive an optimal solution. In order to implement DRL, TensorFlow is employed and a network comprising 10 F-APs and 5 RRHs serving 30 users is simulated. Comparisons with other schemes relying on weighted minimum mean square error (WMMSE), Q-learning, fixed, and random resource allocation reveal that the proposed DRL algorithm achieves improved convergence while latency is reduced by 18\% to 49\% compared to the other schemes. Meanwhile, the cache hit rate of DRL outperforms that of LFU, LRU and FIFO since DRL achieves 89\% hit rate, LFU provides 81\% hit rate while LRU and FIFO provide 75\% hit rate.

In an F-RAN comprising a cloud-based BBU and multiple
cache-aided enhanced RRHs (eRRHs), the authors in \cite{moon2019spl} target the delivery latency minimization over X-haul links when the statistical properties of file popularity are time-varying and unknown. The proposed model-free RL-based scheme relies on linear value function approximation and adaptively activates the backhaul or the fronthaul at each transmission period. Backhaul activation updates the content at the caches of the eRRHs and reduces the latency at future transmission periods, while fronthaul activation leads to cooperative transmissions and reduces the latency at the current transmission period. Performance evaluation in a topology where the BBU is located at the center of the cell while eRRHs and users are circularly placed shows that for low eRRH cache sizes, i.e., less or equal to 4 files, fronthaul selection guarantees lower latency due to the limited potential of caching. On the contrary, for cache sizes larger than 4 files, backhaul activation is superior. Overall, RL provides the lowest latency compared to other schemes relying on only fronthaul/backhaul selection, greedy fronthaul/backhaul selection and offline caching of the most popular files.

Further results for service delay reduction, were given in \cite{wei2019iotj} where the joint optimization of content caching, computation offloading, and radio resource allocation in fog-enabled IoT was studied. AC-based learning is adopted, relying on DNN for Q-value function value approximation of the critic, while the actor policy is represented by another DNN. In addition, the issue of RL divergence is mitigated by employing fixed target network and experience replay. At the same time, the Natural policy-gradient method is used, being more efficient than Standard policy-gradient and guaranteeing that convergence to the local maximum is avoided \cite{peters2018nc}. Results for a library of 1000 different contents highlight that the proposed DRL solutions offers reduced service delay, for cache sizes below 500 contents, still outperforming a conventional content popularity-based caching strategy for cache sizes larger than 500 and up to 1000 contents where identical performance is observed.

\subsection{Caching efficiency}
The improvement of the cache hit rate performance in F-RANs is the main topic of \cite{lu2019vtc}. Towards this end, a distributed edge caching scheme is developed relying on Q-learning with value function approximation (VFA) for reduced complexity and improved convergence. Since content popularity is considered to be unknown, a content request model based on hidden Markov
process is proposed to identify the characteristics of the varying spatio-temporal traffic requests. At the same time, the distributed learning scheme enables each F-AP to independently determining the optimal caching policy, thus avoiding network coordination overheads. Performance evaluation in a network with 20 F-APs, each having a fixed cache size of 5 files, show that the proposed Q-VFA-learning better adapts to content popularity fluctuations and dynamic user arrivals and departures from the network, compared to LRU, LFU and Q-learning without VFA.

Distributed edge caching with dynamic content recommendation is the topic of \cite{yan2020icc}. The authors aim at determining an efficient joint caching and content recommendation policy to reduce the cost of cache replacement in F-APs when user requests data sets are not available. Thus, a per-user request model is presented to characterize the fluctuation of requests after content recommendation. Next, a DDQN-based caching algorithm is formulated in order to reduce the large state and action spaces and guarantee faster convergence. Simulations reveal that the DDQN-based policy offers the highest net profit compared to LRU, LFU, Q-learning and DQN alternatives, for different cache sizes. 

\subsection{Spectral efficiency}
The improvement of QoS provisioning to users given the fluctuations of user preferences is investigated in \cite{zhou2018iccc}. In order to better exploit the caching resources of F-APs, random linear network coding (RLNC) is used to divide the files into subfiles and distribute them across the F-APs. By exploiting the accumulated user requests and considering the successful transmission probability as the reward for the operation of DRL, the optimal caching strategy is derived. Simulations performed using TensorFlow, reveal that the integration of RLNC into the proposed learning solution can save significant caching resources and increase the successful transmission probability in F-RANs, compared to uncoded caching. 

The reduction of the fronthaul load is the main target of the study in \cite{zhou2019icii}. In order to improve the content placement process with unknown file popularity, a two-phase procedure is proposed. In the first phase, feature extraction is employed to extract the content popularity from the frequently collected user requests. In the second phase, DRL with transfer learning is adopted and exploits the predicted file popularity of the previous phase to determine the optimal content placement strategy. Performance comparisons with traditional NN-based algorithms indicate the unsupervised learning-based popularity prediction scheme improves the prediction accuracy, independently of the time-slot duration, while DRL wih transfer learning outperforms both LRU and LFU, in terms of fronthaul load reduction.

In \cite{sun2019pimrc}, the joint optimization of caching and radio resources is targeted. Following a hierarchical approach, a cloud-based cache resource manager aims at maximizing the system throughput and minimizing the storage cost at the F-APs in the long-term. Meanwhile, F-APs are responsible for RRM in the short-term, considering content placement, channel state information (CSI) and user requests. Moreover, interference mitigation is guaranteed by enabling the F-APs to form clusters and perform joint transmissions to the users. In order to achieve improved performance, multi-agent RL is employed, creating one agent per each file and F-AP pair, jointly learning the caching strategy by utilizing historical CSI and user requests data provided by the network information server. Performance evaluation shows that the efficiency of the multi-agent RL-based resource management clearly surpasses that of other schemes, relying on full caching, no caching and caching with fixed probability.

Joint user association and content placement for network payoff maximization is the topic of \cite{yan2020iotj}, in a two-tier network consisting of a massive multiple-input multiple-output (MIMO) macro BS and a group of F-APs. Here, network payoff is defined as the ergodic rate performance utility minus the fronthaul cost for cache replacement. 
In this setting, game theory is invoked, formulating a hierarchical Stackelberg game where at short time scales, users act as followers, dynamically adjusting their F-AP selection, according to content placement status, while at long time scales, the F-APs act as leaders, updating their caches, based on the user association status and the content popularity prediction of a central unit, located at the core network and storing user request data. Towards providing low-complexity and accurate popularity prediction, a stacked auto-encoder (SAE)-based scheme is adopted. Regarding content placement, a DRL-based algorithm, extending \cite{lillicrap2019arxiv} is developed. The DRL architecture is based on online DQN learning where a greedy algorithm selects an action from the state space and offline deep DNN and replay memory creation, executing specific optimization and storing historical information. Performance evaluation shows that DRL offers an average prediction accuracy of 90\%, while baseline DNN- and CNN-based algorithms achieve 80\% and 70\% accuracy, respectively. Finally, compared to LRU and LFU, DRL yields the highest reward, by better capturing the effect of user requests and the amount of data routed through the fronthaul.

%\textbf{Device-to-device:}
%\cite{8698845}~iRAF 
%solves the complex resource allocation problem for the collaborative mobile edge computing (CoMEC)
%network using a deep neural network to predict resource allocation in a self-supervised learning manner.

%\textbf{UAVs:}
%\cite{9209079}~In this paper they optimize the trajectory of each UAV independently with a multi-agent deep reinforcement learning algorithm (Multi-Agent Deep Deterministic Policy Gradient)

%\textit{Vehicular?}
%\cite{9026935}~The authors investigate the computation offloading scheduling problem in vehicular edge computing. They model the prblem by a carefully designed MDP and resort to DRL to deal with the enormous state space.

%\textbf{Device-to-device:}
%\cite{7427078}~In this paper they use a distributed reinforcement learning algorithm for code offloading to ensure low-latency service delivery. They use MDP to
%model such a dynamic environment.

\begin{table}[!htbp]
\caption{List of works focusing on reinforcement learning-aided caching for Fog RANs.}
\label{fog_table}
\centering
{\begin{tabular}[t]{m{9em}<{\raggedright} m{9em}<{\raggedright} m{9em}<{\raggedright}}
\toprule
\textbf{Reference}	& \textbf{Performance target}  & \textbf{RL solution}\\
\midrule%
\vspace{0.15cm}Guo et al.~\cite{guo2020access} &\vspace{0.15cm} Delay &\vspace{0.15cm} Dueling DQN \\%
\vspace{0.15cm}Rahman et al.~\cite{rahman2020tvt} &\vspace{0.15cm} Delay &\vspace{0.15cm} DQN \\%
\vspace{0.15cm} Moon et al.~\cite{moon2019spl} &\vspace{0.15cm} Delay &\vspace{0.15cm} Model-free RL \\%
\vspace{0.15cm} Wei et al.~\cite{wei2019iotj} &\vspace{0.15cm} Delay &\vspace{0.15cm} AC-based DRL \\%
\vspace{0.15cm} Lu et al.~\cite{lu2019vtc} &\vspace{0.15cm} Cache hit rate &\vspace{0.15cm} Q-VFA-learning \\%
\vspace{0.15cm} Yan et al.~\cite{yan2020icc} &\vspace{0.15cm} Cache replacement cost &\vspace{0.15cm} DDQN \\%
\vspace{0.15cm} Zhou et al.~\cite{zhou2018iccc} &\vspace{0.15cm} Average success rate &\vspace{0.15cm} DRL \\%
\vspace{0.15cm} Zhou et al.~\cite{zhou2019icii} &\vspace{0.15cm} Fronthaul usage &\vspace{0.15cm} DRL with transfer learning \\%
\vspace{0.15cm} Sun et al.~\cite{sun2019pimrc} &\vspace{0.15cm} System throughput, storage costs &\vspace{0.15cm} Multi-agent RL \\%
Yan et al.~\cite{yan2020iotj}  & \vspace{0.15cm} Ergodic rate, fronthaul usage &  \vspace{0.15cm} DRL\\
\bottomrule
\end{tabular}}
\end{table}

\section{Cooperative Networks}\label{coop}

{Cooperation among network nodes has been considered as a viable means for improving the quality of communication by improving the wireless conditions through increased diversity and intelligent transmission scheduling \cite{laneman2004tit,bletsas2007twc, michalopoulos2008twc}. Furthermore, data buffering at edge nodes, in the form of dedicated relays or user devices provide reduced outages and higher data rates \cite{zlatanov2014cm, nomikos2016cst, nomikos2018access, nomikos2021wc}. In the context of edge caching, cooperative schemes can be applied both in content caching in a distributed manner, as well as by employing user devices to cache content of other users.}

\subsection{Cooperative Caching}
%Machine learning has greatly helped in the problem of resource management in edge computing networks to improve the quality of service and experience allowing to model complex caching scenarios that would be difficult model otherwise. \todo{This serves as an introductory paragraph with general content that has been discussed in Section I. I think it can be removed.}

\begin{figure}[t]
\centering
\includegraphics[width=\columnwidth]{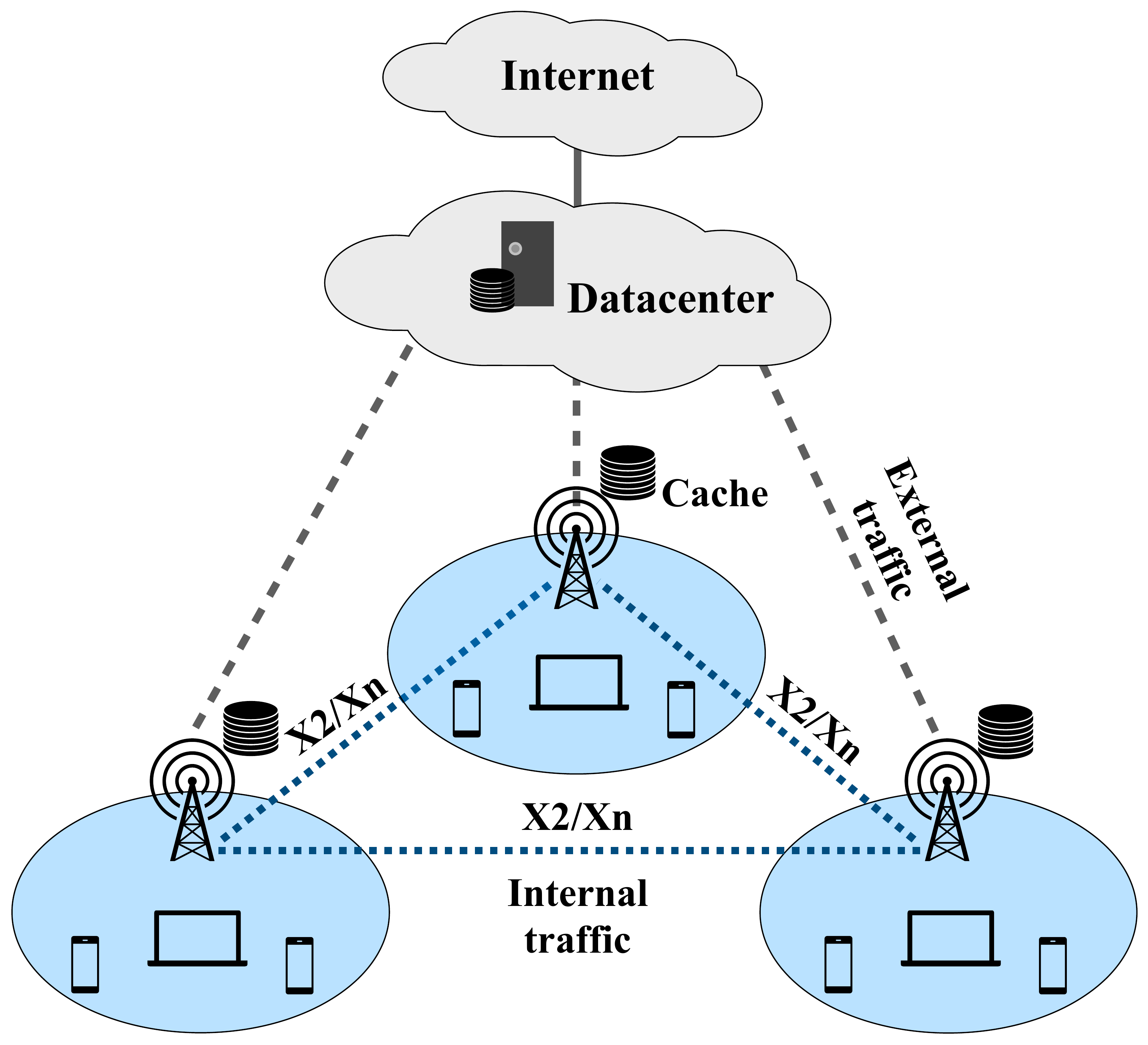}
\caption{A network of cooperating BSs where data from the local cache is shared through the X2/Xn interface and constant data fetching from the core network is avoided.}
\label{coop_caching}
\end{figure}

%Machine learning approaches can be grouped into three main categories: supervised learning, unsupervised learning and reinforcement learning. Supervised learning infers labeling the data and defining a learning function while unsupervised learning focuses on unlabeled or raw data. Both approaches are suitable for offline scenarios while reinforcement learning is considered an online learning algorithm as it continuously monitors response on actions taken and measures against a reward~\cite{sutton1998}. \todo{We discuss ML categories in Section I. We can use this information there and remove it from here.}
A general network architecture comprising cache-aided BSs cooperate and share data through X2/Xn interface in order to avoid constant data fetching from the core network is depicted in Fig.~\ref{coop_caching}. Relevant works are discussed in the following sections, focusing on various performance targets. % In the following sections we present these works and we group them in the aforementioned categories.

\subsubsection{Delay reduction}
%[a]
In~\cite{radenkovic2020access} the authors solve the problem of content caching using a multi-agent AC DRL approach in which edges adaptively learn their best caching policies. {More specifically, in dynamic environments, CognitiveCache is proposed, enabling edge nodes to learn their best caching policies and collaborate with their neighbouring nodes to optimize content placement, thus reducing latency and transmission cost.} RL can provide solutions in volatile and unreliable environments where there is no global knowledge. Single-agent DRL caching has been proposed in~\cite{zhong2018iss, sadeghi2019ccn} where a single edge node has to make the most suitable caching decisions to act on a global network. This requires from every single node to have its own caching policy and a single central agent to make the global decisions resulting in a huge action space~\cite{wang2020infocom}. %The authors of~\cite{radenkovic2020access} focus on a Q-learning actor-critic method which 
{CognitiveCache offers better convergence than~\cite{wang2020infocom} while comparisons with DQNCache \cite{sadeghi2019ccn}, ProbCache \cite{psaras2012icn}, LRU and LFU show that it reduces latency by 33\%, 47\%, 66\%, 71\% and transmission cost by 23\%, 75\%, 83\% and 87\%, respectively.}

A slightly different problem is considered in~\cite{ren2020access} where the capability of users to offload computing tasks to edge computing nodes is examined. Therefore, the coordination between edge computing nodes for the management of the compute and cache resources is investigated. Several challenges have to be addressed in this context like the uncertainty of the computing task, the workload scheduling of a single node but also the resource allocations of multiple nodes during the computation of a collaborative task. Last but not least, low latency of collaborative computing has to be ensured. {A simulation environment has been developed in Python and DDQN is compared against dueling DQN, DQN and Natural Q-learning, in terms of task computation failures, revealing better performance due to improved caching decisions, resulting in reduced delay.}

Single and joint transmission of nodes are considered in~\cite{lin2020tvt} where storage- and transmission-level cooperation is exploited to optimize content caching and updating for video delivery. The authors formulate the problem as an MDP where the reward is mapped to the level of delay reduction. They develop an online RL algorithm to search for the optimal caching policy {updating cache contents in an online manner. In addition, the proposed Q-learning algorithm is extended with linear approximation, thus facilitating its application in settings with a large number of contents. Comparisons in terms of normalized delivery delay are given against two algorithms relying on traditional optimization. The first algorithm has been presented in \cite{chen2017twc} and operates by allocating part of the caches in each cluster to store the most popular content in every edge BS, while the remaining parts cooperatively cache different partitions of the less popular content in different nodes. The second algorithm is the FemtoCaching strategy of \cite{shanmugan2013tit}, employing nodes with low-rate backhaul capacity but large storage to cache popular video content while non-cached files are transmitted by the cellular BS. Results suggest that the proposed strategy provides a delay reduction of at least 6\% and 15\%, compared to the first and second algorithm, respectively.} %\todo{In the rest of the paragraph (I commented it out) you include four works studying content caching but you do not say how they compare to \cite{lin2020tvt}. For example, stating that "5G intelligent networks where studied in \cite{ning2019tis, ning2020tii}" does not give any information regarding RL and edge caching. Up to now, we discuss each work individually and we describe the RL solution and the comparisons with other ML or conventional solutions. I commented it out but if you want to expand the discussion we can re-integrate it :)} % There are various related works that have not well addressed the content caching problem for real-time demands. A reinforcement learning approach for resource allocation is considered in~\cite{sutton1998mit}. 5G intelligent networks were studied in~\cite{ning2019tis, ning2020tii}. A deep learning algorithm for dynamic channel environments was studied in~\cite{guo2019tvt}, where a reinforcement learning algorithm was used for bandwidth allocation and buffer management.

One of the main challenges in DRL is the need of the agent to observe enough features of the environment to ensure decision accuracy. The authors of~\cite{luo2020fwc} propose an AC-based DRL approach for multi-cell and single-cell cooperative networks {where BSs compete with each other for wireless access and also cooperate towards reducing the average delay. In this context, the agents decide their individual caching actions while cooperating with each other, resulting in a centralized critic network and a decentralized actor network. In this framework, the agents update the actor network with their observations and the critic network with the complete state space. Compared to LRU, LFU and FIFO in a scenario with time-varying content popularity, the AC-based DRL offers the best long-term performance and each time the popularity distribution changes, it is able to converge to the previous delay performance level.}%For addressing the problem of edge caching in small base stations and user equipment, content popularity is analyzed. Such analysis is either impractical to be frequently performed because of significant use of resources or assumptions lead to less precise results. A solution to such efficient caching problems is the use of machine learning. 

The decentralized cooperative BS caching problem to minimize the content access latency is the topic of \cite{wang2020iotj}. The proposed solution relies on FL and DRL and its main novelty lies in the fact that it uses two rounds of training. During the first round, the BSs learn a shared predictive model using training parameters as the initial input of local training. In the second round, the BSs upload the near-optimal local parameters as input of the global training. They reduce the performance loss and average delay while they improve the hit rate. More specifically, they compare their proposed platform against baseline schemes like LRU, LFU, FIFO, achieving improved performance. In addition the decentralized cooperative approach performs very close to an alternative centralized DRL algorithm.

The authors of \cite{jiang2019icc, jiang2019access} {propose a MAB-based cooperative caching policy to reduce the download latency in a multi-cell MEC network. The difference of \cite{jiang2019access} to~\cite{jiang2019icc} is that while the user's preference is unknown, the historical content demands are available.} In contrast to other algorithms that make assumptions on the content popularity distribution~\cite{jiang2017tmc}, this work does not depend on previous knowledge of content popularity and user's preference. A Q-learning algorithm on the MEC servers which they train with their own, local, caching decisions and subsequently they combine the results with the decisions of other MEC servers. The authors resolve to a combinatorial MAB upper confidence bound method to reduce the complexity of the problem. {Performance comparisons of the MAB-based algorithm against} a single-agent RL caching algorithm, a modified version of LRU and a randomised replacement caching algorithm {suggest that the weighted average download latency can be reduced by 8\%, 21\% and 24\%, respectively.} %\todo{Merged the two papers and it is OK!}

%[e]
The joint investigation of cooperative edge caching at BSs and request routing, towards reducing the content access delay of the users and improving their QoS is at the epicenter of \cite{li2019wcnc}. This problem is modelled as an MDP and DDQN-based learning is adopted for providing both QoS guarantees and backhaul offloading without statistical knowledge of the content popularity. As a reward, this scheme considers the long-term average content fetching delay of the end-users. Trace-driven simulations suggest that DDQN-based learning offers a performance gain of 7\%, 11\% and 9\%, in terms of delay reduction when compared to LRU, LFU and FIFO caching schemes, respectively. At the same time, the performance gap against an oracle algorithm having a priori knowledge of the users' preferences and behavior is at 4\%.

\subsubsection{Caching efficiency}
{The authors of~\cite{luo2020fwc} also formulate a cache hit rate maximization problem and solve it through AC-based DRL in cooperative topologies. Performance comparisons against LRU, LFU and FIFO for different cache sizes reveal that by employing a centralized critic network, the AC-based DRL learns how the individual decisions of the agents impact the overall cache hit rate, striking a balance between the cache hit rate of each agent and that of the overall system. As a result, the cache hit rate of the learning-based framework surpasses the performance of the three conventional caching policies.}

Then, in \cite{chien2020fgcs}, the authors combine SDN and C-RAN architectures for improved cooperative edge caching. The BBUs of C-RAN are equipped MEC capabilities, resulting in intelligent BBU pools, performing signal processing and data pre-processing and improving the performance of AI applications. In order to maximize the cache hit rate and the cache capacity usage, DRL is employed, considering both global and local cache information. The proposed Q-learning-based collaborative cache algorithm (Q-LCCA) selects among three different actions, regarding cache management. The first action is to cache the most popular data in the local cache of BBUs. The second action exploits neighboring BBUs with short transmission delays, being coordinated by the SDN controller and data in the global cache, formed by the caches of multiple BBUs. As a third action, Q-LCCA adopts random data caching. A roulette-based policy is presented to optimize action selection, applying weights to each one. Performance comparisons, using Matlab$^\copyright$, shows improved cache hit rate, as the number of content types increases, over alternative schemes relying only on local and global caching.%\todo{It needs extension in order to clarify what is the cooperative aspect, how the RL solutions operate and how they perform.}

%Several works have focused on using deep reinforcement learning algorithms which better adapt in a volatile environment. The authors of~\cite{sadeghi2018stsp} implemented a Q-learning algorithm to find the optimal caching policy, while in~\cite{zhong2018iss} and~\cite{wei2019iotj} the  authors use an actor-critic deep reinforcement learning framework for caching. Multi-agent reinforcement learning solutions were studied by~\cite{sung2016icmla} and~\cite{jiang2018infocom} while multi-armed bandit-based caching schemes were presented by~\cite{song2017ccswn} and~\cite{sengupta2014iswcs}. \todo{\cite{zhong2018iss} has been discussed in the first paragraph of this section while \cite{wei2019iotj} and \cite{sengupta2014iswcs} have been already discussed in a previous section. Also, judging from its title \cite{jiang2018infocom} refers to D2D networks and should go to the next subsection while \cite{sadeghi2018stsp} I am not sure it has to do with cooperative caching. For references \cite{jiang2018infocom} and \cite{song2017ccswn} can you re-write/extend them according to the additions that I made (with \NN{blue}) in this section? \textbf{The general logic is one paragraph for each work containing: one sentence stating the problem, two-three sentences describing the topology and RL solutions and one sentence presenting results and comparisons.} This should be done for the rest of the references in Sections IV and V in order to provide a clear view of each work.}

Wireless content delivery and content replacement are the focus of~\cite{sung2016icmla}. The authors apply multi-agent Q-learning to improve content caching performance, modeling the problem as an MDP, where each small BS is considered as an agent. The caching policy aims at maximizing the cache hit rate and combines LFU and LRU policies, achieving a better performance than the standalone versions. In the simulations. they adopt the shot noise model \cite{traverso2013sigcom} for determining the content request pattern over time, generating content requests with temporal correlation. Results show that under temporal correlation, the learning-based approach reaches a hit rate of 79\%, compared to 77\% and 76\% for LFU and LRU. Meanwhile, delay performance is improved and the multi-agent Q-learning guarantees a delay of 1.1 time-slots, while LFU and LRU provide delays of 1.3 and 1.45 time-slots, respectively. This study is highly related to~\cite{gu2014icc} which also addresses content replacement as an MDP. However, the two works aim at different objectives, since in \cite{sung2016icmla} the goal is to maximize the cache hit rate while the objective of~\cite{gu2014icc} is to minimize the system transmission cost. Moreover the work in \cite{sung2016icmla} adopts more realistic simulation parameters and network procedures, as multiple contents can be simultaneously replaced.

%\subsubsection{Spectral efficiency}

\begin{table}[!htbp]
\caption{List of works focusing on reinforcement learning-aided caching for cooperative networks.}
\label{coop_table}
\centering
{\begin{tabular}[t]{m{9em}<{\raggedright} m{9em}<{\raggedright} m{9em}<{\raggedright}}
\toprule
\textbf{Reference}	& \textbf{Performance target}  & \textbf{RL solution}\\
\midrule%
\vspace{0.15cm}Radenkovic et al.~\cite{radenkovic2020access} &\vspace{0.15cm} Delay, transmission cost &\vspace{0.15cm} AC-based DRL \\%
\vspace{0.15cm}Ren et al.~\cite{ren2020access} &\vspace{0.15cm} Delay &\vspace{0.15cm} DDQN \\%
\vspace{0.15cm}Lin et al.~\cite{lin2020tvt} &\vspace{0.15cm} Delay &\vspace{0.15cm} MDP \\%
\vspace{0.15cm}Luo et al.~\cite{luo2020fwc} &\vspace{0.15cm} Delay, cache hit rate &\vspace{0.15cm} AC-based DRL \\%
\vspace{0.15cm}Wang et al.~\cite{wang2020iotj} &\vspace{0.15cm} Delay &\vspace{0.15cm} FL and DRL \\%
\vspace{0.15cm}Jiang et al.~\cite{jiang2019icc, jiang2019access} &\vspace{0.15cm} Delay &\vspace{0.15cm} MAB-based RL \\%
\vspace{0.15cm}Li et al.~\cite{li2019wcnc} &\vspace{0.15cm} Delay &\vspace{0.15cm} DDQN \\%
\vspace{0.15cm}Chien et al.~\cite{chien2020fgcs} &\vspace{0.15cm} Cache hit rate &\vspace{0.15cm} Q-learning \\%
\vspace{0.15cm}Sung et al.~\cite{sung2016icmla} &\vspace{0.15cm} Cache hit rate &\vspace{0.15cm} Multi-agent Q-learning \\%
\vspace{0.15cm}Gu et al.~\cite{gu2014icc} &\vspace{0.15cm} Transmission cost &\vspace{0.15cm} Q-learning \\%
\vspace{0.15cm}Tang et al.~\cite{tang2020tii} &\vspace{0.15cm} Energy consumption &\vspace{0.15cm} Q-learning and DQN \\%
\vspace{0.15cm}Yin et al.~\cite{yin2018gsip} &\vspace{0.15cm} Energy consumption, delay &\vspace{0.15cm} DQN \\%
\vspace{0.15cm}Jiang et al.~\cite{jiang2018infocom} &\vspace{0.15cm} Delay &\vspace{0.15cm} Combinatorial MAB-based RL \\%
\vspace{0.15cm}Zhang et al.~\cite{zhang2020network, zhang2020twc} &\vspace{0.15cm} Traffic offloading &\vspace{0.15cm} DQN \\%
\vspace{0.15cm}He et al.~\cite{he2018gc,he2018wc, he2020tnse} &\vspace{0.15cm} Backhaul usage reduction, utility reduction &\vspace{0.15cm} DRL \\%
%\vspace{0.1cm}He et al.~\cite{he2018wc, he2020tnse} &\vspace{0.1cm} Backhaul usage reduction, utility reduction &\vspace{0.1cm} DQL \\%
\vspace{0.15cm}Wang et al.~\cite{wang2019access} &\vspace{0.15cm} System throughput &\vspace{0.15cm} Policy-gradient DRL \\%
Sun et al.~\cite{sun2018jnca} &\vspace{0.15cm} Offloading ratio &\vspace{0.15cm} Combinatorial MAB-based RL%
\\
\bottomrule
\end{tabular}}
\end{table}

\subsection{Device-to-Device Networks}
The massive connectivity requirements of 6G networks necessitate the integration of novel communication paradigms, deviating from conventional architectures. Thus, D2D communication has been proposed as a remedy for excessive cellular traffic, enabling users to directly cooperate and exchange data or perform relaying for cell edge users. Caching at the users devices and intelligent D2D resource allocation can provide several gains to wireless networks, minimizing among others, energy consumption, delay and backhaul usage \cite{penda2017gc, chen2017cl, lin2019vtm}.

\subsubsection{Energy efficiency}

%[d]
In~\cite{tang2020tii}, the authors aim at improving caching efficiency in D2D networks and minimize the energy cost by prefetching the optimal files at user's devices and small BSs. D2D communication is used to offload part of the cellular traffic, exploiting the caching capabilities of users and their distribution. The request behaviours are modeled as an MDP and RL is applied to discover the file popularity and user preferences. Because of the different capabilities and algorithm complexities, a Q-learning algorithm is applied on users' devices and DQN on small BSs. Comparisons against optimal caching with known popularity, random caching file selection and the case without user caching, depict that the proposed learning-based algorithm closely follows the performance of optimal caching independently of the number of files and user preferences profiles with the other two schemes significantly falling behind both in terms of energy consumption and cache hit rate.

%[f]
The work in ~\cite{yin2018gsip} focuses on both content placement
and delivery strategies in cache-enabled D2D networks, aiming at minimizing the content delivery delay and the power consumption. For this purpose, ESN-based learning is employed for predicting the content popularity and user mobility patterns, determining what and where to cache. Then, content delivery is optimized by relying on a DQN-based algorithm, exploiting CSI and content transmission delay observations to decide which actions should be taken. The DQN-based caching strategy is evaluated in a multi-user network and compared against Q-learning and random caching. It is observed that due to the larger action-state space, the reward in the DQN case is higher than that of Q-learning, while random caching provides significantly smaller rewards.

%[c]
\subsubsection{Delay reduction}

Apart from \cite{yin2018gsip}, jointly tackling energy and delay performance concerns through a two-step learning process, i.e., ESN-based prediction and DQN-based content delivery, there have been various RL strategies focusing on delay reduction.

Learning-based caching strategies are proposed in~\cite{jiang2018infocom} for D2D caching where multi-agent MAB modeling is adopted. Since the action space is too large, there is no instantaneous knowledge of the content popularity profile. More specifically, the users follow a Q-learning approach where each one learns the Q-values via their own actions, as well as considering the actions of other users. In order to reduce the action space and the overall complexity, a belief-based modified combinatorial UCB approach is adopted for regret minimization, in terms of download latency. They conduct simulation experiments to compare the performance of the proposed algorithm against conventional caching schemes, including random replacement, LRU and LFU. The proposed algorithm outperforms the baseline algorithms in average download latency and cache hit rate and its advantage increases as the cache unit size increases, while offering the best performance independently of the number of files. %\todo{Rephrased it, extended it and it's OK!}

%[a, b]
\subsubsection{Spectral efficiency}

The authors in \cite{zhang2020network, zhang2020twc}~address D2D mobile edge caching for traffic offloading. They present a content delivery market formulation and their solution is based on blockchains and smart contracts. More specifically, the caching problem includes a content placement and cache sharing problem, as well as the verification that the caching actions of the peers are recorded and handled in a trustworthy manner. In this topology, there exist different subsystems performing necessary procedures, integrated by a cache and blockchain controller. First, the caching subsystem associates the peer contributions with their willingness to share data via D2D communication. In addition, the blockchain subsystem is responsible for transaction verification which should happen at low cost and latency, while ensuring system scalability. Both problems are formulated as an MDP and are addressed by a DQN algorithm. Comparisons of the proposed DQN algorithm against the DQL approach of \cite{he2018wc,he2020tnse} and a greedy scheme, employing each node to cache the most popular content within its coverage show improved offloading performance and highlight the importance of nodes' participation in data sharing through efficient incentive mechanisms.

%[e]
Another dimension that can be exploited towards optimal resource allocation in D2D-aided mobile edge computing networks is related to social relationships among users, as presented in~\cite{he2018gc,he2018wc,he2020tnse}. %The authors do not make any assumptions or simplifications and they perform a deep Q-learning algorithm to solve the resource allocation problem.The authors of~\cite{he2018gc} elaborate the problem of in-network caching and device-to-device communications in mobile social networks.
Here, the targeted reward is mapped to increasing the MNO's profits through improved backhaul usage for video content delivery, considering the knowledge of the social relationships, in terms of trust among users. The social trust scheme that they present uses Bayesian inference for direct observation through past observations and Dempster-Shafer for indirect observations, combining evidence from multiple users' belief \cite{chen2005ic}. Towards solving the resource allocation problem, a big data DQL resource allocation strategy to make optimal decisions for network resources and these decisions are based on network observations and not on any explicit control rules. Simulations are presented using TensorFlow, as well as comparisons with a scheme without indirect observation, a scheme without edge computing \cite{bu2010milcom} and a scheme without D2D communications \cite{su2015cm}. It is observed that DRL increases the backhaul usage benefiting the profits of MNOs, independently of the number of content types, while the total utility performance is better than that of the other schemes for different numbers of malicious D2D transmitters.

%[g]
In D2D-aided information-centric wireless networks, collaborative caching at the users' devices can result in improved spectral efficiency and offloading. Thus, the authors in~\cite{wang2019access} study resource allocation and power control in a small-cell MEC network with D2D communication. Initially, depending on whether or not the cache of a user is not empty, a selection among cellular or D2D communication is performed. When D2D communication occurs, channel reuse increases and an efficient power allocation scheme should be devised. So, 
a policy-gradient DRL approach is presented to select the power levels, using the Gaussian distribution, as well as softmax channel selection for maximizing the spectral efficiency and minimizing the interference. Comparisons with DQN reveal improved spectral efficiency and reduced interference, since the proposed solution carries out in a continuous state space, while DQN selects among discrete power levels.

%[h][i]
%Mobile social networks is the topic of~\cite{he2018wc} and~\cite{he2020tnse} where the authors study the impact of social relationships on mobile edge computing and device-to-device communications. They propose a deep reinforcement learning approach where each agent receives from its peers measurements and observations. This information combined with the wireless channel conditions and the trust value of every peer are given as input to a deep neural network which outputs proposed actions. Based on the action performed there is an operator revenue which is observed and given back to the agent as a reward. The latter is also used to train and optimise the neural network. The distinct features of this work are the social trust scheme which improves the security of mobile social networks and the fact that they use Bayesian inference approach and Dempster-Shafer theory for the evaluation of direct and indirect observations. The proposed deep reinforcement learning approach is implemented using Google TensorFlow and simulation results demonstrate algorithm convergence as well as its effectiveness on the reduction of the backhaul usage. They also study the performance of the trust scheme and the impact of malicious transmitters on the network performance.

\subsubsection{Caching efficiency}

%[j]
The work in \cite{sun2018jnca} identified two challenging issues that need to be addressed in D2D networks. First, the need to make caching decisions towards maximizing the probability that the requested content will be cached in a neighbouring node given the storage constraints and the plethora of available content. Second, the level of traffic offloading when multiple helper nodes are able to cache and deliver the desired content to another user. To address these issues, a caching strategy considering parameters, such as the predicted content popularity, user preferences, user activity level, and social relationships is proposed. %Several works have focused on user preference [16][17] or social relationship [13] but they lack of a comprehensive user profile which is a mixture of different characteristics like user preferences, activity degrees and relationships.
%Moreover, a metric, namely the expected correlation coefficient (ECC) to characterise the relationships among users is presented. 
For characterizing the offloading potential of users, the expected
correlation coefficient is introduced, capturing the offloading probability and the offloading gain obtained
after a D2D content delivery event. At the same time, D2D user pairs are formed by relying on an online learning algorithm, based on the combinatorial MAB framework. Comparisons with random caching, the least frequently caching and matching algorithm of \cite{shafiq2014acm} and the collaborative caching and auction matching algorithm of \cite{jiang2016jsac} are presented, using real-world traces from the Unical/SocialBlueconn dataset to model the D2D topology. Results in terms of offloading due to D2D communication indicate that the MAB-based algorithm achieves a 53\% offloading ratio, contrary to 45\% by collaborative caching and auction matching, 22\% by least frequently caching and matching and 10\% by random caching.

\section{Vehicular Networks}\label{veh}
\begin{figure}[t]
\centering
\includegraphics[width=\columnwidth]{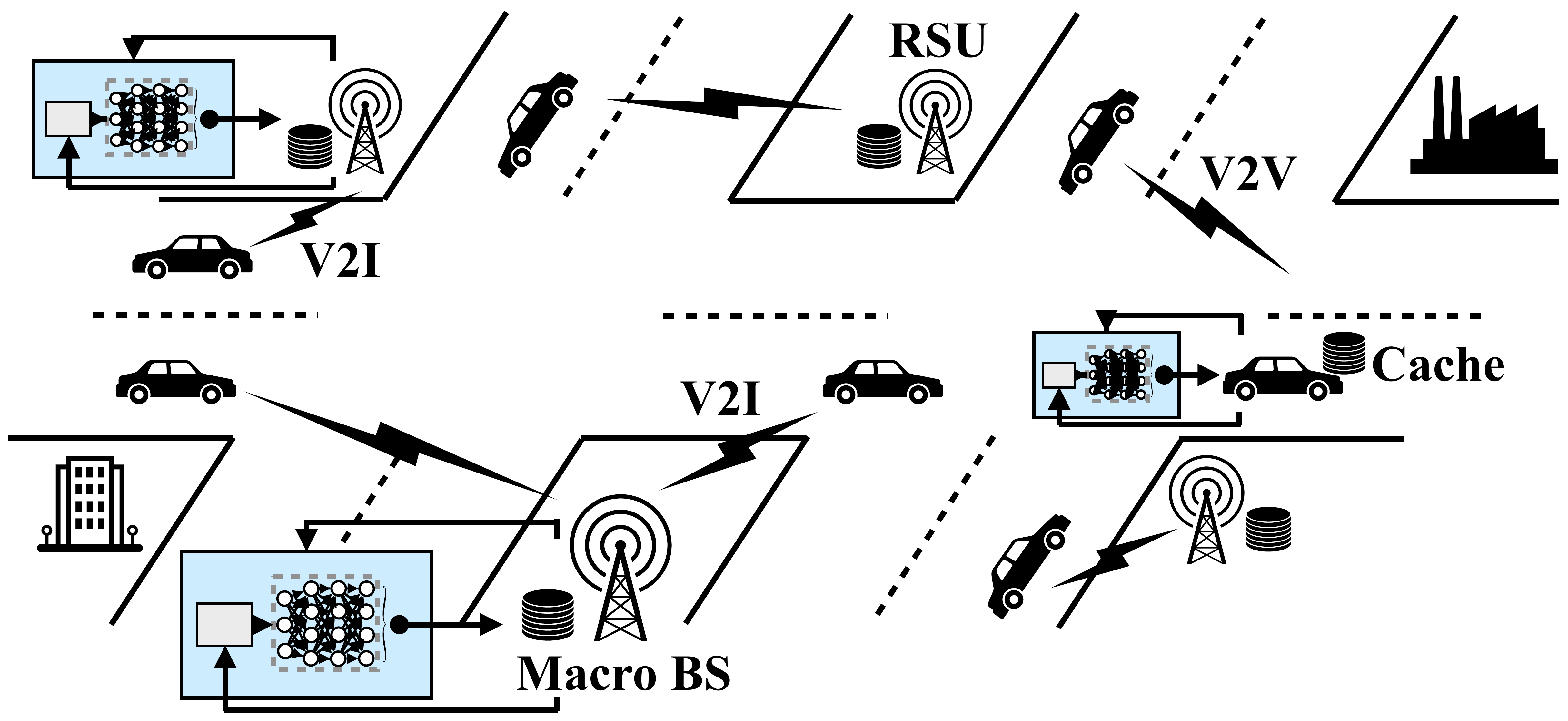}
\caption{Various cases of vehicular edge caching relying on RL-aided nodes.}
\label{v2v}
\end{figure}

Edge caching provides both opportunities and challenges for vehicular networks, related to the dynamicity of the network, the computing and caching power of vehicles and road side units (RSUs), as well as the amount of data generated by participating vehicles. Different scenarios of vehicular communications, such as V2V or vehicle-to-infrastructure (V2I), either towards a macro BS or an RSU are shown in Fig.~\ref{v2v} where RL-aided edge caching is performed. The works presented in this section cover various aspects of these networks, aiming at energy efficiency, delay reduction and improved caching performance, as well as optimizing computation and communication-related metrics. % In the following sections we present these works and we group them in the aforementioned categories.

\subsection{Energy efficiency}
%[a]
In~\cite{dai2019wc} the authors propose a learning framework, allowing cross-layer offloading and cooperative multi-point caching. In cross-layer offloading, a computationally heavy task can be offloaded to the next computation layer, such as RSUs and the latter can subsequently offload it, if necessary, to BSs. Thus, a resource allocation scheme, based on the DDPG RL algorithm is proposed, maximizing the system utility which considers the energy consumption, as well as computation and caching improvement. Simulations demonstrate the effectiveness of the proposed scheme focusing on the cumulative reward of the vehicular network.

The study of \cite{dai2019wcsp, dai2020tvt} aims at solving a two-fold problem: (a) Establishing a network for secure content caching and (b) ensuring efficient content caching in a volatile network. To address (a), the authors propose a blockchain enabled distributed content caching network allowing the base stations and the vehicles to establish a secure peer-to-peer transaction environment. The base stations maintain the permissioned blockchain and the vehicles perform the content caching. To address (b), they propose an optimal content caching scheme based on DRL, taking mobility into account while they accelerate block verification by a new selection method that they propose. The reward function considers the total consumed energy for content transmission and content caching. Also, a security analysis is conducted, showing that the proposed blockchain content caching provides security and privacy protection with low-energy consumption. Performance analysis using a real dataset shows that the proposed algorithm achieves a 86\% of successful content caching requests against 76\% of a greedy algorithm and 5\% of a random content caching algorithm.
%[j]
%Furthermore, the work in \cite{dai2019wcsp} elaborates on the topic of content caching in a vehicular network, without neglecting the security perspective. Towards this end, the authors propose a secure V2V content caching scheme, based on permissioned blockchains, ensuring privacy where the BSs maintain the permissioned blockchain. In order to withstand the volatile environment, DRL is employed, coupled with block verifiers to accelerate block verification. The performance of the algorithm is compared against two baseline algorithms (greedy and random), showing higher average reward which is expressed, in terms of energy energy consumption for V2V content delivery.

\subsection{Delay reduction}
%[e]
The topic of~\cite{dai2019iccc} is to minimize the content delivery in vehicular edge networks. The authors propose a framework where the RSUs collaborate with vehicles to cache popular contents. Multiple ways are provided to the vehicles to retrieve the needed content, given their limited cache: via (a) V2V links, vehicle-to-RSUs and (c) vehicle-to-macro BSs. They follow a DDPG DRL approach to model the content edge caching and delivery problem where the goal is to minimize the total content delivery latency. The performance is compared against a random edge caching and delivery scheme and a DDPG scheme without bandwidth optimization. The proposed algorithm demonstrated significant improvements against the other two approaches in terms of total content delivery latency and cumulative total reward, while exhibiting faster convergence.

%[d]
The authors of~\cite{qiao2020iotj} focus on content delivery in vehicular edge networks, optimizing the content placement and content delivery by taking into account trilateral collaborations among vehicles, macro BSs and RSUs. The problem is modeled as a double time scale MDP (DTS-MDP), considering that vehicle placement and network changes are more frequent than content changes in time. The content placement initially relies on content popularity, vehicle path and resource availability. These are also the conditions to optimize in the large time-scale, while in the small time-scale, vehicle scheduling and bandwidth allocation are performed, targeting the minimization of content delivery latency. In this context, a DDPG framework is adopted for obtaining a sub-optimal solution with low computational complexity. Performance evaluation is presented, taking into account content delivery latency, content hit-ratio and system cost where substantial improvements against a random caching scheme and a non-cooperative caching scheme (e.g. 135.62\% and 34.33\% content hit ratio increase accordingly) are shown.

%[b]
The authors of~\cite{luo2020iotj} elaborate on the problem of vehicular edge computing in which they take into consideration RSUs able to provide computation offloading and data bandwidth. Moreover, vehicles can collaborate using V2V communications with data relays and collaborative computing. Targeting to improve the data processing performance, a penalty mechanism is introduced, dictating imposing penalties if the data processing deadline is not satisfied. To ensure the above, in~\cite{luo2020iotj}, they propose a framework to model communication, computation and caching. Then, they propose a DQN algorithm to find the optimal strategy for a collaborative data scheduling scheme which will minimize the system-wide data processing cost with ensured application delays. They perform benchmarks where they demonstrate data processing cost reduction and help data to be processed under delay constraints.

\subsection{Spectrum efficiency}
%[f]
The paper in \cite{he2020tvt} focuses on management of content caching, computing and networking at vehicular networks. They propose a DRL algorithm that uses a DQN for the approximation of a Q-value action. The proposed integrated framework orchestrates networking, caching and computing resources to meet the requirements of different applications. The reward function considers the MVNO revenue, consisting of the received SNR of the wireless access
link, the task computation capability and the cache state. The authors present simulations results of their proposed approach against schemes without virtualization, MEC offloading and mobile edge caching in a setting where the system state is assumed to be static. The total reward of this work depends on computation, communication and caching. The proposed results show the superiority of the proposed approach in terms of total utility.

%[g]
Another scheme focusing on maximizing the MVNO revenue is presented in \cite{chen2020cc}, proposing dynamic resource allocation in vehicle networks. The reward function include the MVNO revenue, which is modeled as a function of the received SNR, the computation capability and the access link state. By relying on SDN and the principles of information-centric networking, dynamic orchestration of computing and communication resources for virtual wireless network optimization is targeted. The authors model the resource allocation strategy as an MDP, using function approximation. The high complexity of the problem is tackled using A3C-based RL. Simulation results demonstrate increased reward and good convergence speed, resulting in improved MVNO revenue compared to a conventional scheme without learning capabilities.

The authors of~\cite{ning2020tits} take advantage of DRL to orchestrate edge computing and resource allocation with the goal of maximizing the MNO revenue without degrading the users' QoE in V2V networks. More specifically, they design a DDPG model to optimize the problem of resource allocation and task assignment in a volatile vehicular environment with mobile edge caching servers. They conduct experiments based on real traffic data and they compare their proposed algorithm against a non-cooperative scheme, a computation offloading scheme and an edge caching scheme without computation offloading. They demonstrate that the MNO profits can be significantly larger when the proposed scheme is adopted, compared to the other benchmark schemes.

%[i]
In~\cite{tan2018tvt} the goal is to maximize the reward of a vehicular network, considering the combined reward of communication, computation and data offloading, while satisfying a deadline-constrained service. In the proposed network, both vehicles and RSUs have caching storage and computing capabilities. They collaborate and communicate via D2D communication to achieve cache hits for the vehicles when content is retrieved via nearby vehicles or RSUs. Moreover, vehicles are able to offload tasks to neighbouring vehicles, RSUs and, in the lack of available resource at those, to BSs. A Q-learning algorithm with multi-timescale network is proposed for the caching placement, computing resource allocation and assessment of the sets of possible connecting RSUs and vehicles. They use optimal parameter configuration for their proposed algorithm to validate their theoretical findings and to shown significant cost gains against a random resource allocation scheme and a scheme where caching and computing capabilities are limited to RSUs.

%[k]
A similar topic with \cite{tan2018tvt} is studied  in \cite{tan2019tvt}. The authors formulate the problem as a joint caching and computing allocation problem for cost minimization under the constraints of dynamic storage capacities of RSUs. They propose multi-time scale algorithms for caching placement, computing resource allocation and assessment of the sets of potentially connecting RSUs, contrary to connecting RSUs and vehicles, as  in~\cite{tan2018tvt}. The developed algorithms are based on particle swarm optimization and DQL for large and small timescale models, respectively. Numerical results show significant performance gains while using optimal parameter configurations for the proposed algorithms.

%\subsection{Caching efficiency}
%[c]
%Vehicular edge caching enables vehicles that are located in proximity to store large amounts of data. However, the intrinsic nature and the volatility of such networks, as well as security concerns regarding the caching location of sensitive personal data raise new challenges.

\begin{table}[!htbp]
\caption{List of works focusing on reinforcement learning-aided caching for vehicular networks.}
\label{vehicular_table}
\centering
{\begin{tabular}[t]{m{9em}<{\raggedright} m{9em}<{\raggedright} m{9em}<{\raggedright}}
\toprule
\textbf{Reference}	& \textbf{Performance target}  & \textbf{RL solution}\\
\midrule%
\vspace{0.15cm}Dai et al.~\cite{dai2019wc} &\vspace{0.15cm} Energy consumption &\vspace{0.15cm} DDPG-based \\%
\vspace{0.15cm}Dai et al.~\cite{dai2019wcsp,dai2020tvt} &\vspace{0.15cm} Energy consumption &\vspace{0.15cm} DRL and permissioned blockchain \\%
\vspace{0.15cm}Dai et al.~\cite{dai2019iccc} &\vspace{0.15cm} Delay &\vspace{0.15cm} DDPG-based \\%
\vspace{0.15cm}Qiao et al.~\cite{qiao2020iotj} &\vspace{0.15cm} Delay &\vspace{0.15cm} DDPG-based \\%
\vspace{0.15cm}Luo et al.~\cite{luo2020iotj} &\vspace{0.15cm} Computation delay &\vspace{0.15cm} DQN \\%
%[f]
\vspace{0.15cm}He et al.~\cite{he2020tvt} &\vspace{0.15cm} Backhaul usage, MVNO revenue &\vspace{0.15cm} DQN \\%
%[i]
\vspace{0.15cm}Chen et al.~\cite{chen2020cc} &\vspace{0.15cm} Backhaul usage, MVNO revenue &\vspace{0.15cm} A3C-based \\%
\vspace{0.15cm}Ning et al.~\cite{ning2020tits} &\vspace{0.15cm} MNO revenue &\vspace{0.15cm} DRL \\%
\vspace{0.15cm}Tan et al.~\cite{tan2018tvt} &\vspace{0.15cm} Spectral efficiency, storage cost &\vspace{0.15cm} DQL with multi-timescale network \\%
\vspace{0.15cm}Tan et al.~\cite{tan2019tvt} &\vspace{0.15cm} Spectral efficiency, storage cost &\vspace{0.15cm} Particle swarm optimization and DQL %\\\vspace{0.15cm}%
%\vspace{0.1cm}Dai et al.~\cite{dai2019wcsp} &\vspace{0.1cm} Energy efficiency &\vspace{0.1cm} DRL with block verifiers \\\vspace{0.15cm}%
\\
\bottomrule
\end{tabular}}
\end{table}

\section{UAV-Aided Networks}\label{uavs}
UAVs will play a vital role towards improving the performance of 6G wireless networks. UAV-aided networks offer flexibility in network deployment, allocating resource where and when needed, as well as fast recovery after disasters and network outages. However, the increased degrees of freedom will also pose new challenges in mobile edge networks. In this context, ML is expected to provide solutions for improving the efficiency edge caching by deploying UAVs at optimal locations and determining their trajectory and communication parameters.

\subsection{Delay reduction}
Examining the role of cache-enabled UAVs, the paper in \cite{dai2020pc} examined proactive caching for reduced latency and backhaul load. So, multi-objective optimization is targeted for defining parameters, such as minimum number of deployed UAVs, transmit power, UAV-user association, and cache location. In order to solve the multi-objective optimization, RL is adopted for user grouping, performing local search according to the optimal UAV deployment over each group. From the results, the efficiency of the RL method is shown, effectively minimizing the number of required UAVs, compared to the case without caching, as well as their 3-D placement in the network, resulting in improved cache placement.

Aiming to optimize the performance of cache-enabled UAV-aided NOMA networks, the authors in \cite{zhang2020tvt} turn to RL for tackling the dynamic characteristics of UAV movement and content request variations. Initially, long-term optimization of cache placement, user scheduling and power allocation for NOMA is formulated, minimizing the long-term sum delay of ground users. Then, the optimization problem is converted into an MDP and Q-learning is invoked to reach a near-optimal solution. Still, in large-scale networks, Q-learning fails to cope with the large MDP state and action spaces and thus, function approximation-based caching and resource allocation is proposed. Performance evaluation in a multi-cell environment focuses on one cache-enabled and UAV-aided cell. Content delivery delay and cache hit ratio comparisons include the two RL solutions, a greedy algorithm obtaining the optimal delivery delay of the current state, a fixed algorithm caching the popular contents of previous states, employing round robin-based scheduling, as well as fixed power allocation and finally, random content caching and resource allocation. It is concluded that in small-scale networks, Q-learning provides a small performance gap compared to the greedy algorithm, while in large-scale networks, function approximation outperforms both random and fixed algorithms. 

In a UAV-aided small cell topology, supporting virtual reality (VR) applications with stringent delay constraints, content caching and transmission are studied in \cite{chen2019tcom}. Here, UAVs alleviate the burden of backhaul and access links by collecting the contents that users request and transmitting them to the cache-aided small BSs, communicating with the VR users. The joint optimization of caching and transmission is solved by developing a DL algorithm, relying on LSM NNs and ESNs, namely echo liquid state machine (ELSM) DL, identifying the relationship among actions, selection policy of small BSs and user reliability. Compared to conventional DRL, LSM-based RL offers increased prediction accuracy, using historical data while ESNs reduce the training complexity by adjusting only its output weight matrix and avoiding the calculation of the gradients of all the neurons. Simulations are conducted, comparing ELSM with LSM, Q-learning \cite{bennis2010gc} and ESN-based learning \cite{chen2018tcom}. It is shown that ELSM provides a 10\% and 18.4\% reliability improvement against ESN and Q-learning when 35 users exist in the network. In addition, ELSM converges 11.8\% faster, compared to LSM in a network with 11 SBSs.

An LTE cloud network operating in licensed and unlicensed bands is the main focus in \cite{chen2017gc}, studying resource allocation for cache-enabled UAVs supporting ground users. The performance target here is to maintain queue stability, directly affecting the content transmission delay in the network. Constrained by the limited capacity of the UAV-cloud links, LSM is employed to help the UAVs perform content caching and resource management. LSM enables the cloud to efficiently learn user-centric information regarding content request distribution and to facilitate spectrum allocation by the UAVs. This method is extended in \cite{chen2019twc}, where an optimization problem is formulated to maximize the number of users with stable queues. As a solution to this problem, a self-organizing, decentralized algorithm is developed and LSM is employed for joint caching and resource allocation over both licensed and unlicensed bands. Then, performance evaluation is conducted illustrating the increase in the number of users with stable queues in comparison to Q-learning with and without content caching. 

\subsection{Energy efficiency}

Improved UAV-aided network operation in the context of the Internet of Vehicles is the topic of \cite{alhilo2020tits}. Due to the increased mobility and dynamic environment in terms of content requests, content delivery becomes challenging. More specifically, vehicles request contents from the UAV in the downlink while the latter has to decide which popular contents should be cached from arriving vehicles in the uplink. As a performance metric, the maximization of the number of served vehicles over the UAV energy consumption is investigated and the problem of joint caching, UAV trajectory and RRM are tackled. However, the complex environment comprising randomly arriving vehicles makes the use of traditional optimization approaches prohibitive. On the contrary, after formulating the joint problem as an MDP, PPO-based DRL is employed to control UAV trajectory. Here its operation relies on rewarding the agent when a vehicle is served by the UAV, and penalizing the agent according to the energy consumption incurred by moving the UAV. 
Simulations consider a single cache-aided UAV topology and PPO is compared against stationary UAV, random UAV mobility, maximum speed selection for moving the UAV back-and-forth over the highway, as well as minimum energy selection, selecting the UAV velocity resulting in minimum energy consumption. It is revealed that the PPO balances the amount of traffic offloading and the energy consumption at the UAV, while better adapting to content requests, as shown for different content popularity values.

Another RL-based solution for improved energy efficiency in cache-enabled UAV-aided networks is presented in \cite{wu2020wcmc}. Focusing on an urban scenario with mobile users, the storage and energy capabilities of the UAV are considered, towards maximizing the sum achievable throughput. Since this problem is shown to be non-convex, DRL is adopted for joint content placement and trajectory design. More specifically, energy-efficient UAV control is achieved by employing a DDQN for online trajectory design, according to real-time user mobility and avoiding the over-estimation of the value function of traditional DQN. Also, during the offline content placement stage, a link-based caching strategy is developed for cache hit rate maximization through approximation and convex optimization, leading to a trade-off among file popularity and diversity. In order to illustrate the performance of caching and DDQN trajectory design, a multi-cell simulation environment using TensorFlow is developed, showing that the selection of appropriate hyperparameters, such as learning rate can increase the performance of DDQN. As a result, throughput and energy consumption gains are harvested, compared to static circular trajectory designs, as mobile users are not optimally served.

\begin{table}[!htbp]
\caption{List of works focusing on reinforcement learning-aided caching for UAV-aided networks.}
\label{uav_table}
\centering
{\begin{tabular}[t]{m{9em}<{\raggedright} m{9em}<{\raggedright} m{9em}<{\raggedright}}
\toprule
\textbf{Reference}	& \textbf{Performance target}  & \textbf{RL solution}\\
\midrule%
\vspace{0.15cm}Dai et al.~\cite{dai2020pc} &\vspace{0.15cm} Delay, backhaul usage &\vspace{0.15cm} Local search-based \\%
\vspace{0.15cm}Zhang et al.~\cite{zhang2020tvt} &\vspace{0.15cm} Delay &\vspace{0.15cm} Q-learning and function approximation \\%
\vspace{0.15cm}Chen et al.~\cite{chen2019tcom} &\vspace{0.15cm} Delay &\vspace{0.15cm} LSM and ESN-based \\%
\vspace{0.15cm}Chen et al.~\cite{chen2017gc, chen2019twc} &\vspace{0.15cm} No. of stable queue users &\vspace{0.15cm} LSM-based \\%
\vspace{0.15cm}Al-Hilo et al.~\cite{alhilo2020tits} &\vspace{0.15cm} Spectral efficiency, energy consumption &\vspace{0.15cm} PPO \\%
Wu et al.~\cite{wu2020wcmc}  & \vspace{0.15cm} Spectral efficiency, energy consumption &  \vspace{0.15cm} DDQN\\

\bottomrule
\end{tabular}}
\end{table}

\section{Open Issues}\label{open}

\subsection{Physical-layer aspects}

In recent years, RL %\todo{[Why first ML and then RL?]}
has been proposed as an alternative approach to conventional optimization when the derivation of optimal communication parameters entails excessive complexity and high network coordination overheads. In mobile edge networks where different cells might overlap, the design of RL-aided caching policies should not neglect physical-layer issues. These include intra- and inter-cell interference, fast fading due to mobility from receivers and transmitters, as well as path-loss.  The RL solution should evaluate data related to signal-to-interference plus noise ratio and jointly determine the edge caching locations, BS and user association, D2D cooperation, duplexing method, as well as modulation order, coding rate, {beamforming vectors} and transmit power level. RL-aided solutions in the form of MAB have already shown promising performance in such physical-layer-related problems \cite{maghsudi2016cm, nomikos2020mlsp,ghoorchian2021tccn, nomikos2021icc, petropulu2020icassp, petropulu2020tcom}. Also, in many cases, edge networks comprise nodes with limited capabilities, such as IoT devices, requiring high energy efficiency. So, the integration of wireless powered communications with RL-aided edge caching and computing should be studied \cite{psomas2020cl}.

Furthermore, the development of RL-aided solutions integrating high diversity data buffering techniques with edge caching represents an attractive research direction. In recent years, buffer-aided techniques in relay and D2D networks have shown tremendous gains in different communication scenarios, increasing the transmission reliability and mitigating the degrading effects of interference and fading \cite{nomikos2018tvt,kim2019adhoc,simoni2016twc,zhang2015cm}. For example, full-duplex (FD) relays can increase the flexibility in edge caching locations by operating in a hybrid fashion, establishing end-to-end communication of users with a BS, providing their desired content. However, at another instance, the FD relay, having already cached this content, can deliver it under more favorable channel conditions. 

\subsection{Non-orthogonal multiple-access}

Edge caching has the potential to improve the performance of mobile networks, resulting in homogeneous QoS and better backhaul/fronthaul offloading. However, the massive number of users and devices in 6G networks calls for NOMA strategies in order to better exploit the wireless resources. Recently, the integration of NOMA in mobile edge networks has been proposed, jointly determining the task allocation, caching location and power allocation for NOMA \cite{ding2018tcom,xiang2019jstsp,pei2020tvt}. %Nonetheless, only a small number of studies exist on learning-based NOMA strategies for edge caching, e.g. \cite{yang2020icc, yang2020twc, zhang2020tvt}.
At the same time, the benefits of NOMA in buffer-aided networks have already been shown in various works and tailor-made caching policies when users are simultaneously served on the same physical resources should be devised \cite{luo2017cl,nomikos2020tvt,xu2020iotj,li2020tcom}. Still, considering the large amount of network parameters, including storage, power level, spectrum and QoS constraints, conventional optimization will often fail in deriving optimal caching policies when NOMA is employed.

\subsection{Radical learning paradigms}

RL operation for improving the performance of edge networks with a massive number of users and IoT devices should aim at avoiding complex and resource-demanding learning solutions while still exploiting the large geographical distribution and heterogeneity of edge nodes. In this context, the incorporation of transfer and federated learning in RL-aided edge networks represents a fertile research area with only a limited number of contributions \cite{wang2019network, zhou2019icii, wang2020iotj}. First, transfer learning is based on initially extracting features, such as file popularity on a base network with a generalized data set. Then, these features are used to facilitate DRL agents at the edge to converge to the optimal edge caching policy, thus minimizing the energy consumption at the edge devices. On the other hand, FL leverages the observations of multiple DRL agents at different edge nodes and trains a shared model. In addition, communication costs among edge nodes are reduced, since FL uses locally stored data, only calculating updates to the global shared model of the coordinating node.

\subsection{Security and trust}

Mobile edge networks comprise operator-owned clouds, infrastructure-based BSs and machines, as well as user devices. {Caching, apart from rate and delay improvements, has the potential to improve security in such heterogeneous wireless networks, e.g., physical-layer security (PLS) \cite{Zhao:2018}. While the problem has been studied extensively in different context, ranging from cellular \cite{ZhengTCOM:2018} to cooperative networks \cite{Petropulu:2010},} cooperative and low-complexity RL solutions can be implemented on a wide range of network nodes {to facilitate and enhance PLS}, {especially when trade-offs among latency and security arise \cite{petropulu2020cl}}. At the same time, issues of trust are raised, especially in infrastructure-less D2D-aided edge scenarios where social-awareness can be exploited \cite{he2018gc, he2018wc,he2020tnse}. RL algorithms should take into consideration the behaviour of cooperating nodes and incur penalties when malicious behavior is observed, since caching at small BSs and more importantly, at user devices can threaten user data privacy. Furthermore, in decentralized learning paradigms, such as federated learning which better suit privacy sensitive applications, it is necessary to ensure that shared models will be based on information exchange among trustworthy peers. In this area, recent works have adopted blockchains and smart contracts, highlighting their efficiency in M2M, D2D and V2V RL-aided edge caching but still, further advancements are needed \cite{li2020iotj, zhang2020network, zhang2020twc, dai2019wcsp, dai2020tvt}.

\subsection{Cooperative caching extensions}
The problem of cooperative caching is examined from various perspectives and different avenues for further research are open. In~\cite{radenkovic2020access}, where the edges adaptively learn their best caching policies using a multi-agent AC DLR, the authors envision to improve their algorithms' accuracy, scalability and efficiency in heterogeneous networks by using real-time heuristics and analytics. In a different kind of edge caching setup, like in~\cite{ren2020access}, where the problem of user offloading tasks to edge computing nodes is examined and the coordination between edge computing nodes for the management of the compute and cache resources is investigated, an area of further research is the usage of competitive bidding and allocation priorities. Furthermore, since they elaborate the problem of workload scheduling but also resource allocation in collaborative tasks, user security on the edge becomes an interesting area for further research and investigation. In~\cite{luo2020fwc}, the BS compete against each other for wireless access and also cooperate towards reducing the average delay. The authors are interested in jointly solving the content caching problem along with other problems related to power control and user scheduling. 

\subsection{Volatile networking topologies}
Vehicular and UAV-aided networks represent interesting domains of edge caching where BSs, RSUs, ground and aerial vehicles collaborate for the optimal management of caching and computation resources in a highly volatile environment. Further research can be performed regarding security, resource management and mobility prediction. More specifically, the authors of~\cite{dai2019wc} take into account RSUs to provide computation offloading and spectrum planning to investigate AI algorithms for efficient handover. %Moreover security and user privacy issues will be investigated.
Proactive caching and pre-allocation of network bandwidth is also the future focus of~\cite{ning2020tits}. Also, the caching and computing resources orchestration for different application types, aiming at increased energy efficiency in highly mobile networks provides another interesting future direction \cite{he2020tvt}.

%\todo{\subsection{Other ideas? Mobile network control - ???}}

%\todo{...}

\section{Conclusions}\label{conclusions}

Edge caching represents a major shift in network architecture design, since content is brought closer to the users in an intelligent and proactive manner. In this way, the burden in backhaul and fronthaul is relieved and repeated requests to remote web servers are avoided. Still, the optimization of edge caching performance must take into consideration several characteristics, including mobility, resource allocation, energy and storage capabilities, as well as {requirements, including} rate and delay. In this context, the adoption of reinforcement learning can lead to tangible performance gains at an acceptable complexity, overcoming the limitations of traditional approaches. This survey focused on reinforcement-aided edge caching in a variety of network settings, comprising fixed access points, fog-enabled paradigms, cooperative schemes, as well as aerial and ground vehicles. The discussion of the different learning solutions revealed that the fusion of learning and edge caching can result in significant benefits, independently of the complexity of the wireless environment and surpass the performance of conventional optimization solutions, while guaranteeing service requirements in an online and autonomous fashion. Finally, several open issues in the field have been highlighted, paving the way for further innovations towards realizing the 6G communications vision.

%\section*{References}
\let\oldaddcontentsline\addcontentsline% Store \addcontentsline
\renewcommand{\addcontentsline}[3]{}% Make \addcontentsline a no-op
\bibliography{Paper}

\begin{thebibliography}{100}
\providecommand{\url}[1]{#1}
\csname url@samestyle\endcsname
\providecommand{\newblock}{\relax}
\providecommand{\bibinfo}[2]{#2}
\providecommand{\BIBentrySTDinterwordspacing}{\spaceskip=0pt\relax}
\providecommand{\BIBentryALTinterwordstretchfactor}{4}
\providecommand{\BIBentryALTinterwordspacing}{\spaceskip=\fontdimen2\font plus
\BIBentryALTinterwordstretchfactor\fontdimen3\font minus
  \fontdimen4\font\relax}
\providecommand{\BIBforeignlanguage}[2]{{%
\expandafter\ifx\csname l@#1\endcsname\relax
\typeout{** WARNING: IEEEtran.bst: No hyphenation pattern has been}%
\typeout{** loaded for the language `#1'. Using the pattern for}%
\typeout{** the default language instead.}%
\else
\language=\csname l@#1\endcsname
\fi
#2}}
\providecommand{\BIBdecl}{\relax}
\BIBdecl

\bibitem{itu2015}
ITU, \emph{IMT traffic estimates for the years 2020 to 2030}, 2015.

\bibitem{barbarossa2014spm}
S.~{Barbarossa}, S.~{Sardellitti}, and P.~{Di Lorenzo}, ``Communicating while
  computing: Distributed mobile cloud computing over 5g heterogeneous
  networks,'' \emph{IEEE Signal Processing Magazine}, vol.~31, no.~6, pp.
  45--55, 2014.

\bibitem{shi2016computer}
W.~{Shi} and S.~{Dustdar}, ``The promise of edge computing,'' \emph{Computer},
  vol.~49, no.~5, pp. 78--81, 2016.

\bibitem{liu2016cm}
D.~{Liu}, B.~{Chen}, C.~{Yang}, and A.~F. {Molisch}, ``Caching at the wireless
  edge: design aspects, challenges, and future directions,'' \emph{IEEE
  Communications Magazine}, vol.~54, no.~9, pp. 22--28, 2016.

\bibitem{zhang2019cst}
C.~{Zhang}, P.~{Patras}, and H.~{Haddadi}, ``Deep learning in mobile and
  wireless networking: A survey,'' \emph{IEEE Communications Surveys
  Tutorials}, vol.~21, no.~3, pp. 2224--2287, 2019.

\bibitem{luo2020fwc}
F.~L. {Luo}, \emph{Deep Multi-Agent Reinforcement Learning for Cooperative Edge
  Caching}, 2020, pp. 439--457.

\bibitem{pan2018iotj}
J.~{Pan} and J.~{McElhannon}, ``Future edge cloud and edge computing for
  internet of things applications,'' \emph{IEEE Internet of Things Journal},
  vol.~5, no.~1, pp. 439--449, 2018.

\bibitem{wubben2014spm}
D.~{Wubben}, P.~{Rost}, J.~S. {Bartelt}, M.~{Lalam}, V.~{Savin},
  M.~{Gorgoglione}, A.~{Dekorsy}, and G.~{Fettweis}, ``Benefits and impact of
  cloud computing on 5g signal processing: Flexible centralization through
  cloud-ran,'' \emph{IEEE Signal Processing Magazine}, vol.~31, no.~6, pp.
  35--44, 2014.

\bibitem{zhou2018cm}
Z.~{Zhou}, S.~{Mumtaz}, K.~M.~S. {Huq}, A.~{Al-Dulaimi}, K.~{Chandra}, and
  J.~{Rodriquez}, ``Cloud miracles: Heterogeneous cloud ran for fair
  coexistence of lte-u and wi-fi in ultra dense 5g networks,'' \emph{IEEE
  Communications Magazine}, vol.~56, no.~6, pp. 64--71, 2018.

\bibitem{tran2017cm}
T.~X. {Tran}, A.~{Hajisami}, P.~{Pandey}, and D.~{Pompili}, ``Collaborative
  mobile edge computing in 5g networks: New paradigms, scenarios, and
  challenges,'' \emph{IEEE Communications Magazine}, vol.~55, no.~4, pp.
  54--61, 2017.

\bibitem{piao2019iotj}
Z.~{Piao}, M.~{Peng}, Y.~{Liu}, and M.~{Daneshmand}, ``Recent advances of edge
  cache in radio access networks for internet of things: Techniques,
  performances, and challenges,'' \emph{IEEE Internet of Things Journal},
  vol.~6, no.~1, pp. 1010--1028, 2019.

\bibitem{zhang2017wcm}
H.~{Zhang}, Y.~{Qiu}, X.~{Chu}, K.~{Long}, and V.~C.~M. {Leung}, ``Fog radio
  access networks: Mobility management, interference mitigation, and resource
  optimization,'' \emph{IEEE Wireless Communications}, vol.~24, no.~6, pp.
  120--127, 2017.

\bibitem{wu2018wcm}
D.~{Wu}, L.~{Zhou}, Y.~{Cai}, and Y.~{Qian}, ``Collaborative caching and
  matching for d2d content sharing,'' \emph{IEEE Wireless Communications},
  vol.~25, no.~3, pp. 43--49, 2018.

\bibitem{gurugopinath2020network}
S.~{Gurugopinath}, Y.~{Al-Hammadi}, P.~C. {Sofotasios}, S.~{Muhaidat}, and
  O.~A. {Dobre}, ``Non-orthogonal multiple access with wireless caching for
  5g-enabled vehicular networks,'' \emph{IEEE Network}, vol.~34, no.~5, pp.
  127--133, 2020.

\bibitem{zhao2019vtm}
N.~{Zhao}, F.~R. {Yu}, L.~{Fan}, Y.~{Chen}, J.~{Tang}, A.~{Nallanathan}, and
  V.~C.~M. {Leung}, ``Caching unmanned aerial vehicle-enabled small-cell
  networks: Employing energy-efficient methods that store and retrieve popular
  content,'' \emph{IEEE Vehicular Technology Magazine}, vol.~14, no.~1, pp.
  71--79, 2019.

\bibitem{samuel1959ibm}
A.~L. {Samuel}, ``Some studies in machine learning using the game of
  checkers,'' \emph{IBM Journal of Research and Development}, vol.~3, no.~3,
  pp. 210--229, 1959.

\bibitem{alpaydin2010mit}
E.~Alpaydin, \emph{Introduction to Machine Learning}, 2nd~ed.\hskip 1em plus
  0.5em minus 0.4em\relax The MIT Press, 2010.

\bibitem{mcmahan2017aistats}
H.~McMahan, E.~Moore, D.~Ramage, S.~Hampson, and B.~A. y~Arcas,
  ``Communication-efficient learning of deep networks from decentralized
  data,'' in \emph{2017 International Conference on Artificial Intelligence and
  Statistics (AISTATS)}, 2017, pp. 1--10.

\bibitem{gunduz2019jsac}
D.~{Gunduz}, P.~{de Kerret}, N.~D. {Sidiropoulos}, D.~{Gesbert}, C.~R.
  {Murthy}, and M.~{van der Schaar}, ``Machine learning in the air,''
  \emph{IEEE Journal on Selected Areas in Communications}, vol.~37, no.~10, pp.
  2184--2199, 2019.

\bibitem{li2018cst}
L.~{Li}, G.~{Zhao}, and R.~S. {Blum}, ``A survey of caching techniques in
  cellular networks: Research issues and challenges in content placement and
  delivery strategies,'' \emph{IEEE Communications Surveys Tutorials}, vol.~20,
  no.~3, pp. 1710--1732, 2018.

\bibitem{yao2019cst}
J.~{Yao}, T.~{Han}, and N.~{Ansari}, ``On mobile edge caching,'' \emph{IEEE
  Communications Surveys Tutorials}, vol.~21, no.~3, pp. 2525--2553, 2019.

\bibitem{mcclellan2020as}
M.~McClellan, C.~Cervello-Pastor, and S.~Sallent, ``Deep learning at the mobile
  edge: Opportunities for 5g networks,'' \emph{Applied Sciences}, vol.~10,
  no.~14, 2020.

\bibitem{zhu2018network}
H.~{Zhu}, Y.~{Cao}, W.~{Wang}, T.~{Jiang}, and S.~{Jin}, ``Deep reinforcement
  learning for mobile edge caching: Review, new features, and open issues,''
  \emph{IEEE Network}, vol.~32, no.~6, pp. 50--57, 2018.

\bibitem{anokye2019zte}
S.~Anokye, M.~Seid, and S.~Guolin, ``A survey on machine learning based
  proactive caching,'' \emph{ZTE Communications}, vol.~17, no.~4, pp. 46--55,
  2019.

\bibitem{wang2020informatics}
Y.~Wang and V.~Friderikos, ``A survey of deep learning for data caching in edge
  network,'' \emph{Informatics}, vol.~7, no.~4, 2020.

\bibitem{chen2019cst}
M.~{Chen}, U.~{Challita}, W.~{Saad}, C.~{Yin}, and M.~{Debbah}, ``Artificial
  neural networks-based machine learning for wireless networks: A tutorial,''
  \emph{IEEE Communications Surveys \& Tutorials}, vol.~21, no.~4, pp.
  3039--3071, 2019.

\bibitem{sheraz2020comst}
M.~Sheraz, M.~Ahmed, X.~Hou, Y.~Li, D.~Jin, Z.~Han, and T.~Jiang, ``Artificial
  intelligence for wireless caching: Schemes, performance, and challenges,''
  \emph{IEEE Communications Surveys Tutorials}, vol.~23, no.~1, pp. 631--661,
  2021.

\bibitem{yang2019icc}
Z.~{Yang}, Y.~{Liu}, Y.~{Chen}, and G.~{Tyson}, ``Deep reinforcement learning
  in cache-aided mec networks,'' in \emph{ICC 2019 - 2019 IEEE International
  Conference on Communications (ICC)}, 2019, pp. 1--6.

\bibitem{yang2020icc}
Z.~{Yang}, Y.~{Liu}, and Y.~{Chen}, ``Distributed reinforcement learning for
  noma-enabled mobile edge computing,'' in \emph{2020 IEEE International
  Conference on Communications Workshops (ICC Workshops)}, 2020, pp. 1--6.

\bibitem{yang2020twc}
Z.~{Yang}, Y.~{Liu}, Y.~{Chen}, and N.~{Al-Dhahir}, ``Cache-aided noma mobile
  edge computing: A reinforcement learning approach,'' \emph{IEEE Transactions
  on Wireless Communications}, vol.~19, no.~10, pp. 6899--6915, 2020.

\bibitem{sadeghi2018icassp}
A.~{Sadeghi}, F.~{Sheikholeslami}, A.~G. {Matrques}, and G.~B. {Giannakis},
  ``Reinforcement learning for 5g caching with dynamic cost,'' in \emph{2018
  IEEE International Conference on Acoustics, Speech and Signal Processing
  (ICASSP)}, 2018, pp. 6653--6657.

\bibitem{sadegh2019jsac}
A.~{Sadeghi}, F.~{Sheikholeslami}, A.~G. {Marques}, and G.~B. {Giannakis},
  ``Reinforcement learning for adaptive caching with dynamic storage pricing,''
  \emph{IEEE Journal on Selected Areas in Communications}, vol.~37, no.~10, pp.
  2267--2281, 2019.

\bibitem{zhong2020tccn}
C.~{Zhong}, M.~C. {Gursoy}, and S.~{Velipasalar}, ``Deep reinforcement
  learning-based edge caching in wireless networks,'' \emph{IEEE Transactions
  on Cognitive Communications and Networking}, vol.~6, no.~1, pp. 48--61, 2020.

\bibitem{lillicrap2019arxiv}
T.~P. Lillicrap, J.~J. Hunt, A.~Pritzel, N.~Heess, T.~Erez, Y.~Tassa,
  D.~Silver, and D.~Wierstra, ``Continuous control with deep reinforcement
  learning,'' \emph{CoRR}, vol. abs/1509.02971, 2019.

\bibitem{dulac2015arxiv}
G.~Dulac{-}Arnold, R.~Evans, P.~Sunehag, and B.~Coppin, ``Reinforcement
  learning in large discrete action spaces,'' \emph{CoRR}, vol. abs/1512.07679,
  2015.

\bibitem{wu2019cl}
P.~{Wu}, J.~{Li}, L.~{Shi}, M.~{Ding}, K.~{Cai}, and F.~{Yang}, ``Dynamic
  content update for wireless edge caching via deep reinforcement learning,''
  \emph{IEEE Communications Letters}, vol.~23, no.~10, pp. 1773--1777, 2019.

\bibitem{mnihnature2015}
V.~Mnih, K.~Kavukcuoglu, D.~Silver, A.~Rusu, and \textit{et al.}, ``Human-level
  control through deep reinforcement learning,'' \emph{Nature}, vol. 518, no.
  7540, pp. 529--533, 2015.

\bibitem{wu2019access}
W.~{Wu}, Y.~{Gao}, T.~{Zhou}, Y.~{Jia}, H.~{Zhang}, T.~{Wei}, and Y.~{Sun},
  ``Deep reinforcement learning-based video quality selection and radio bearer
  control for mobile edge computing supported short video applications,''
  \emph{IEEE Access}, vol.~7, pp. 181\,740--181\,749, 2019.

\bibitem{he2019cim}
X.~{He}, K.~{Wang}, and W.~{Xu}, ``Qoe-driven content-centric caching with deep
  reinforcement learning in edge-enabled iot,'' \emph{IEEE Computational
  Intelligence Magazine}, vol.~14, no.~4, pp. 12--20, 2019.

\bibitem{ma2020icc}
M.~{Ma} and V.~W.~S. {Wong}, ``A deep reinforcement learning approach for
  dynamic contents caching in hetnets,'' in \emph{ICC 2020 - 2020 IEEE
  International Conference on Communications (ICC)}, 2020, pp. 1--6.

\bibitem{ma2020twc}
------, ``Age of information driven cache content update scheduling for dynamic
  contents in heterogeneous networks,'' \emph{IEEE Transactions on Wireless
  Communications}, pp. 1--1, 2020.

\bibitem{yates2017tit}
Y.~Sun, E.~Uysal-Biyikoglu, R.~D. Yates, C.~E. Koksal, and N.~B. Shroff,
  ``Update or wait: How to keep your data fresh,'' \emph{IEEE Transactions on
  Information Theory}, vol.~63, no.~11, pp. 7492--7508, 2017.

\bibitem{yates2021cst}
R.~D. Yates, Y.~Sun, D.~R. Brown, S.~K. Kaul, E.~Modiano, and S.~Ulukus, ``Age
  of information: An introduction and survey,'' \emph{IEEE Journal on Selected
  Areas in Communications}, vol.~39, no.~5, pp. 1183--1210, 2021.

\bibitem{yatesisit2017}
R.~D. {Yates}, P.~{Ciblat}, A.~{Yener}, and M.~{Wigger}, ``Age-optimal
  constrained cache updating,'' in \emph{2017 IEEE International Symposium on
  Information Theory (ISIT)}, 2017, pp. 141--145.

\bibitem{zhang2019iccc}
Z.~{Zhang} and M.~{Tao}, ``Accelerated deep reinforcement learning for wireless
  coded caching,'' in \emph{2019 IEEE/CIC International Conference on
  Communications in China (ICCC)}, 2019, pp. 249--254.

\bibitem{xu2020twc}
X.~{Xu}, M.~{Tao}, and C.~{Shen}, ``Collaborative multi-agent multi-armed
  bandit learning for small-cell caching,'' \emph{IEEE Transactions on Wireless
  Communications}, vol.~19, no.~4, pp. 2570--2585, 2020.

\bibitem{xu2018gc}
X.~{Xu} and M.~{Tao}, ``Collaborative multi-agent reinforcement learning of
  caching optimization in small-cell networks,'' in \emph{2018 IEEE Global
  Communications Conference (GLOBECOM)}, 2018, pp. 1--6.

\bibitem{wei2018icc}
Y.~{Wei}, Z.~{Zhang}, F.~R. {Yu}, and Z.~{Han}, ``Joint user scheduling and
  content caching strategy for mobile edge networks using deep reinforcement
  learning,'' in \emph{2018 IEEE International Conference on Communications
  Workshops (ICC Workshops)}, 2018, pp. 1--6.

\bibitem{li2020iotj}
M.~{Li}, F.~R. {Yu}, P.~{Si}, W.~{Wu}, and Y.~{Zhang}, ``Resource optimization
  for delay-tolerant data in blockchain-enabled iot with edge computing: A deep
  reinforcement learning approach,'' \emph{IEEE Internet of Things Journal},
  vol.~7, no.~10, pp. 9399--9412, 2020.

\bibitem{liu2019tii}
M.~{Liu}, F.~R. {Yu}, Y.~{Teng}, V.~C.~M. {Leung}, and M.~{Song}, ``Performance
  optimization for blockchain-enabled industrial internet of things (iiot)
  systems: A deep reinforcement learning approach,'' \emph{IEEE Transactions on
  Industrial Informatics}, vol.~15, no.~6, pp. 3559--3570, 2019.

\bibitem{zhang2020tcom}
X.~{Zhang}, G.~{Zheng}, S.~{Lambotharan}, M.~R. {Nakhai}, and K.~{Wong}, ``A
  reinforcement learning-based user-assisted caching strategy for dynamic
  content library in small cell networks,'' \emph{IEEE Transactions on
  Communications}, vol.~68, no.~6, pp. 3627--3639, 2020.

\bibitem{sengupta2014iswcs}
A.~{Sengupta}, S.~{Amuru}, R.~{Tandon}, R.~M. {Buehrer}, and T.~C. {Clancy},
  ``Learning distributed caching strategies in small cell networks,'' in
  \emph{2014 11th International Symposium on Wireless Communications Systems
  (ISWCS)}, 2014, pp. 917--921.

\bibitem{thar2019access}
K.~{Thar}, T.~Z. {Oo}, Y.~K. {Tun}, D.~H. {Kim}, K.~T. {Kim}, and C.~S. {Hong},
  ``A deep learning model generation framework for virtualized multi-access
  edge cache management,'' \emph{IEEE Access}, vol.~7, pp. 62\,734--62\,749,
  2019.

\bibitem{thar2019cnsm}
K.~{Thar}, T.~Z. {Oo}, Z.~{Han}, and C.~S. {Hong}, ``Meta-learning-based deep
  learning model deployment scheme for edge caching,'' in \emph{2019 15th
  International Conference on Network and Service Management (CNSM)}, 2019, pp.
  1--6.

\bibitem{tensorflow}
\BIBentryALTinterwordspacing
Tensorflow, (accessed December, 2020). [Online]. Available:
  \url{https://www.tensorflow.org}
\BIBentrySTDinterwordspacing

\bibitem{keras}
\BIBentryALTinterwordspacing
Keras, (accessed December, 2020). [Online]. Available: \url{https://keras.io}
\BIBentrySTDinterwordspacing

\bibitem{liu2019access}
D.~{Liu} and C.~{Yang}, ``A deep reinforcement learning approach to proactive
  content pushing and recommendation for mobile users,'' \emph{IEEE Access},
  vol.~7, pp. 83\,120--83\,136, 2019.

\bibitem{wangicml2016}
Z.~Wang, T.~Schaul, M.~Hessel, H.~Hasselt, M.~Lanctot, and N.~Freitas,
  ``Dueling network architectures for deep reinforcement learning,'' in
  \emph{Proceedings of The 33rd International Conference on Machine Learning},
  ser. Proceedings of Machine Learning Research, M.~F. Balcan and K.~Q.
  Weinberger, Eds., vol.~48.\hskip 1em plus 0.5em minus 0.4em\relax New York,
  New York, USA: PMLR, 20--22 Jun 2016, pp. 1995--2003.

\bibitem{mnih2016icml}
V.~Mnih, A.~P. Badia, M.~Mirza, A.~Graves, T.~Lillicrap, T.~Harley, D.~Silver,
  and K.~Kavukcuoglu, ``Asynchronous methods for deep reinforcement learning,''
  in \emph{Proceedings of The 33rd International Conference on Machine
  Learning}, ser. Proceedings of Machine Learning Research, M.~F. Balcan and
  K.~Q. Weinberger, Eds., vol.~48.\hskip 1em plus 0.5em minus 0.4em\relax New
  York, New York, USA: PMLR, 20--22 Jun 2016, pp. 1928--1937.

\bibitem{schulman2017arxiv}
J.~Schulman, F.~Wolski, P.~Dhariwal, A.~Radford, and O.~Klimov, ``Proximal
  policy optimization algorithms,'' \emph{CoRR}, vol. abs/1707.06347, 2017.

\bibitem{wang2019network}
X.~{Wang}, Y.~{Han}, C.~{Wang}, Q.~{Zhao}, X.~{Chen}, and M.~{Chen}, ``In-edge
  ai: Intelligentizing mobile edge computing, caching and communication by
  federated learning,'' \emph{IEEE Network}, vol.~33, no.~5, pp. 156--165,
  2019.

\bibitem{guo2019access}
K.~{Guo} and C.~{Yang}, ``Temporal-spatial recommendation for caching at base
  stations via deep reinforcement learning,'' \emph{IEEE Access}, vol.~7, pp.
  58\,519--58\,532, 2019.

\bibitem{fang2019vtc}
Y.~{Fang}, J.~{Xiong}, P.~{Cheng}, and W.~{Zhang}, ``Distributed caching
  popular services by using deep q-learning in converged networks,'' in
  \emph{2019 IEEE 90th Vehicular Technology Conference (VTC2019-Fall)}, 2019,
  pp. 1--5.

\bibitem{zhang2018bmsb}
W.~{Zhang}, J.~{Xiong}, L.~{Gui}, B.~{Liu}, M.~{Qiu}, and Z.~{Shi}, ``On
  popular services pushing and distributed caching in converged overlay
  networks,'' in \emph{2018 IEEE International Symposium on Broadband
  Multimedia Systems and Broadcasting (BMSB)}, 2018, pp. 1--6.

\bibitem{cheng2019tcom}
P.~{Cheng}, C.~{Ma}, M.~{Ding}, Y.~{Hu}, Z.~{Lin}, Y.~{Li}, and B.~{Vucetic},
  ``Localized small cell caching: A machine learning approach based on rating
  data,'' \emph{IEEE Transactions on Communications}, vol.~67, no.~2, pp.
  1663--1676, 2019.

\bibitem{garg2019icassp}
N.~{Garg}, M.~{Sellathurai}, and T.~{Ratnarajah}, ``Content placement learning
  for success probability maximization in wireless edge caching networks,'' in
  \emph{ICASSP 2019 - 2019 IEEE International Conference on Acoustics, Speech
  and Signal Processing (ICASSP)}, 2019, pp. 3092--3096.

\bibitem{qian2020jsac}
Y.~{Qian}, R.~{Wang}, J.~{Wu}, B.~{Tan}, and H.~{Ren}, ``Reinforcement
  learning-based optimal computing and caching in mobile edge network,''
  \emph{IEEE Journal on Selected Areas in Communications}, vol.~38, no.~10, pp.
  2343--2355, 2020.

\bibitem{he2017cm}
Y.~{He}, F.~R. {Yu}, N.~{Zhao}, V.~C.~M. {Leung}, and H.~{Yin},
  ``Software-defined networks with mobile edge computing and caching for smart
  cities: A big data deep reinforcement learning approach,'' \emph{IEEE
  Communications Magazine}, vol.~55, no.~12, pp. 31--37, 2017.

\bibitem{chou2020icc}
P.~{Chou}, W.~{Chen}, C.~{Wang}, R.~{Hwang}, and W.~{Chen}, ``Deep
  reinforcement learning for mec streaming with joint user association and
  resource management,'' in \emph{ICC 2020 - 2020 IEEE International Conference
  on Communications (ICC)}, 2020, pp. 1--7.

\bibitem{Chen2020icnc}
W.~{Chen}, P.~{Chou}, C.~{Wang}, R.~{Hwang}, and W.~{Chen}, ``Live video
  streaming with joint user association and caching placement in mobile edge
  computing,'' in \emph{2020 International Conference on Computing, Networking
  and Communications (ICNC)}, 2020, pp. 796--801.

\bibitem{dai2016icassp}
B.~{Dai} and W.~{Yu}, ``Joint user association and content placement for
  cache-enabled wireless access networks,'' in \emph{2016 IEEE International
  Conference on Acoustics, Speech and Signal Processing (ICASSP)}, 2016, pp.
  3521--3525.

\bibitem{luo2020twc}
J.~{Luo}, F.~R. {Yu}, Q.~{Chen}, and L.~{Tang}, ``Adaptive video streaming with
  edge caching and video transcoding over software-defined mobile networks: A
  deep reinforcement learning approach,'' \emph{IEEE Transactions on Wireless
  Communications}, vol.~19, no.~3, pp. 1577--1592, 2020.

\bibitem{wang2020infocom}
F.~{Wang}, F.~{Wang}, J.~{Liu}, R.~{Shea}, and L.~{Sun}, ``Intelligent video
  caching at network edge: A multi-agent deep reinforcement learning
  approach,'' in \emph{IEEE INFOCOM 2020 - IEEE Conference on Computer
  Communications}, 2020, pp. 2499--2508.

\bibitem{jiang2019twc}
W.~{Jiang}, G.~{Feng}, S.~{Qin}, T.~S.~P. {Yum}, and G.~{Cao}, ``Multi-agent
  reinforcement learning for efficient content caching in mobile d2d
  networks,'' \emph{IEEE Transactions on Wireless Communications}, vol.~18,
  no.~3, pp. 1610--1622, 2019.

\bibitem{guo2020access}
B.~{Guo}, X.~{Zhang}, Q.~{Sheng}, and H.~{Yang}, ``Dueling deep-q-network based
  delay-aware cache update policy for mobile users in fog radio access
  networks,'' \emph{IEEE Access}, vol.~8, pp. 7131--7141, 2020.

\bibitem{rahman2020tvt}
G.~M.~S. {Rahman}, M.~{Peng}, S.~{Yan}, and T.~{Dang}, ``Learning based joint
  cache and power allocation in fog radio access networks,'' \emph{IEEE
  Transactions on Vehicular Technology}, vol.~69, no.~4, pp. 4401--4411, 2020.

\bibitem{moon2019spl}
J.~{Moon}, O.~{Simeone}, S.~{Park}, and I.~{Lee}, ``Online reinforcement
  learning of x-haul content delivery mode in fog radio access networks,''
  \emph{IEEE Signal Processing Letters}, vol.~26, no.~10, pp. 1451--1455, 2019.

\bibitem{wei2019iotj}
Y.~{Wei}, F.~R. {Yu}, M.~{Song}, and Z.~{Han}, ``Joint optimization of caching,
  computing, and radio resources for fog-enabled iot using natural actor-critic
  deep reinforcement learning,'' \emph{IEEE Internet of Things Journal},
  vol.~6, no.~2, pp. 2061--2073, 2019.

\bibitem{peters2018nc}
J.~Peters and S.~Schaal, ``Natural actor-critic,'' \emph{Neurocomputing},
  vol.~71, no.~7, pp. 1180--1190, 2008.

\bibitem{lu2019vtc}
L.~{Lu}, Y.~{Jiang}, M.~{Bennis}, Z.~{Ding}, F.~{Zheng}, and X.~{You},
  ``Distributed edge caching via reinforcement learning in fog radio access
  networks,'' in \emph{2019 IEEE 89th Vehicular Technology Conference
  (VTC2019-Spring)}, 2019, pp. 1--6.

\bibitem{yan2020icc}
J.~{Yan}, Y.~{Jiang}, F.~{Zheng}, F.~R. {Yu}, X.~{Gao}, and X.~{You},
  ``Distributed edge caching with content recommendation in fog-rans via deep
  reinforcement learning,'' in \emph{2020 IEEE International Conference on
  Communications Workshops (ICC Workshops)}, 2020, pp. 1--6.

\bibitem{zhou2018iccc}
Y.~{Zhou}, M.~{Peng}, S.~{Yan}, and Y.~{Sun}, ``Deep reinforcement learning
  based coded caching scheme in fog radio access networks,'' in \emph{2018
  IEEE/CIC International Conference on Communications in China (ICCC
  Workshops)}, 2018, pp. 309--313.

\bibitem{zhou2019icii}
Y.~{Zhou}, S.~{Yan}, and M.~{Peng}, ``Content placement with unknown popularity
  in fog radio access networks,'' in \emph{2019 IEEE International Conference
  on Industrial Internet (ICII)}, 2019, pp. 361--367.

\bibitem{sun2019pimrc}
Y.~{Sun} and M.~{Peng}, ``Joint cache and radio resource management in fog
  radio access networks: A hierarchical two-timescale optimization
  perspective,'' in \emph{2019 IEEE 30th Annual International Symposium on
  Personal, Indoor and Mobile Radio Communications (PIMRC)}, 2019, pp. 1--6.

\bibitem{yan2020iotj}
S.~{Yan}, M.~{Jiao}, Y.~{Zhou}, M.~{Peng}, and M.~{Daneshmand},
  ``Machine-learning approach for user association and content placement in fog
  radio access networks,'' \emph{IEEE Internet of Things Journal}, vol.~7,
  no.~10, pp. 9413--9425, 2020.

\bibitem{laneman2004tit}
J.~N. {Laneman}, D.~N.~C. {Tse}, and G.~W. {Wornell}, ``Cooperative diversity
  in wireless networks: Efficient protocols and outage behavior,'' \emph{IEEE
  Transactions on Information Theory}, vol.~50, no.~12, pp. 3062--3080, 2004.

\bibitem{bletsas2007twc}
A.~{Bletsas}, H.~{Shin}, and M.~Z. {Win}, ``Cooperative communications with
  outage-optimal opportunistic relaying,'' \emph{IEEE Transactions on Wireless
  Communications}, vol.~6, no.~9, pp. 3450--3460, 2007.

\bibitem{michalopoulos2008twc}
D.~S. {Michalopoulos} and G.~K. {Karagiannidis}, ``Performance analysis of
  single relay selection in rayleigh fading,'' \emph{IEEE Transactions on
  Wireless Communications}, vol.~7, no.~10, pp. 3718--3724, 2008.

\bibitem{zlatanov2014cm}
N.~{Zlatanov}, A.~{Ikhlef}, T.~{Islam}, and R.~{Schober}, ``Buffer-aided
  cooperative communications: opportunities and challenges,'' \emph{IEEE
  Communications Magazine}, vol.~52, no.~4, pp. 146--153, 2014.

\bibitem{nomikos2016cst}
N.~{Nomikos}, T.~{Charalambous}, I.~{Krikidis}, D.~N. {Skoutas},
  D.~{Vouyioukas}, M.~{Johansson}, and C.~{Skianis}, ``A survey on buffer-aided
  relay selection,'' \emph{IEEE Communications Surveys Tutorials}, vol.~18,
  no.~2, pp. 1073--1097, 2016.

\bibitem{nomikos2018access}
N.~{Nomikos}, D.~{Poulimeneas}, T.~{Charalambous}, I.~{Krikidis},
  D.~{Vouyioukas}, and M.~{Johansson}, ``Delay- and diversity-aware
  buffer-aided relay selection policies in cooperative networks,'' \emph{IEEE
  Access}, vol.~6, pp. 73\,531--73\,547, 2018.

\bibitem{nomikos2021wc}
N.~{Nomikos}, T.~{Charalambous}, D.~{Vouyioukas}, and G.~K. {Karagiannidis},
  ``When buffer-aided relaying meets full duplex and noma,'' \emph{IEEE
  Wireless Communications}, vol.~28, no.~1, pp. 68--73, 2021.

\bibitem{radenkovic2020access}
M.~{Radenkovic} and V.~S.~H. {Huynh}, ``Cognitive caching at the edges for
  mobile social community networks: A multi-agent deep reinforcement learning
  approach,'' \emph{IEEE Access}, vol.~8, pp. 179\,561--179\,574, 2020.

\bibitem{zhong2018iss}
C.~{Zhong}, M.~C. {Gursoy}, and S.~{Velipasalar}, ``A deep reinforcement
  learning-based framework for content caching,'' in \emph{2018 52nd Annual
  Conference on Information Sciences and Systems (CISS)}, 2018, pp. 1--6.

\bibitem{sadeghi2019ccn}
A.~{Sadeghi}, G.~{Wang}, and G.~B. {Giannakis}, ``Deep reinforcement learning
  for adaptive caching in hierarchical content delivery networks,'' \emph{IEEE
  Transactions on Cognitive Communications and Networking}, vol.~5, no.~4, pp.
  1024--1033, 2019.

\bibitem{psaras2012icn}
I.~Psaras, W.~K. Chai, and G.~Pavlou, ``Probabilistic in-network caching for
  information-centric networks,'' in \emph{Proceedings of the Second Edition of
  the ICN Workshop on Information-Centric Networking}, ser. ICN '12.\hskip 1em
  plus 0.5em minus 0.4em\relax New York, NY, USA: Association for Computing
  Machinery, 2012, pp. 55--60.

\bibitem{ren2020access}
J.~{Ren}, H.~{Wang}, T.~{Hou}, S.~{Zheng}, and C.~{Tang}, ``Collaborative edge
  computing and caching with deep reinforcement learning decision agents,''
  \emph{IEEE Access}, vol.~8, pp. 120\,604--120\,612, 2020.

\bibitem{lin2020tvt}
P.~{Lin}, Q.~{Song}, J.~{Song}, A.~{Jamalipour}, and F.~R. {Yu}, ``Cooperative
  caching and transmission in comp-integrated cellular networks using
  reinforcement learning,'' \emph{IEEE Transactions on Vehicular Technology},
  vol.~69, no.~5, pp. 5508--5520, 2020.

\bibitem{chen2017twc}
Z.~{Chen}, J.~{Lee}, T.~Q.~S. {Quek}, and M.~{Kountouris}, ``Cooperative
  caching and transmission design in cluster-centric small cell networks,''
  \emph{IEEE Transactions on Wireless Communications}, vol.~16, no.~5, pp.
  3401--3415, 2017.

\bibitem{shanmugan2013tit}
K.~{Shanmugam}, N.~{Golrezaei}, A.~G. {Dimakis}, A.~F. {Molisch}, and
  G.~{Caire}, ``Femtocaching: Wireless content delivery through distributed
  caching helpers,'' \emph{IEEE Transactions on Information Theory}, vol.~59,
  no.~12, pp. 8402--8413, 2013.

\bibitem{wang2020iotj}
X.~{Wang}, C.~{Wang}, X.~{Li}, V.~C.~M. {Leung}, and T.~{Taleb}, ``Federated
  deep reinforcement learning for internet of things with decentralized
  cooperative edge caching,'' \emph{IEEE Internet of Things Journal}, vol.~7,
  no.~10, pp. 9441--9455, 2020.

\bibitem{jiang2019icc}
W.~{Jiang}, G.~{Feng}, S.~{Qin}, and Y.~{Liang}, ``Learning-based cooperative
  content caching policy for mobile edge computing,'' in \emph{ICC 2019 - 2019
  IEEE International Conference on Communications (ICC)}, 2019, pp. 1--6.

\bibitem{jiang2019access}
W.~{Jiang}, G.~{Feng}, S.~{Qin}, and Y.~{Liu}, ``Multi-agent reinforcement
  learning based cooperative content caching for mobile edge networks,''
  \emph{IEEE Access}, vol.~7, pp. 61\,856--61\,867, 2019.

\bibitem{jiang2017tmc}
W.~{Jiang}, G.~{Feng}, and S.~{Qin}, ``Optimal cooperative content caching and
  delivery policy for heterogeneous cellular networks,'' \emph{IEEE
  Transactions on Mobile Computing}, vol.~16, no.~5, pp. 1382--1393, 2017.

\bibitem{li2019wcnc}
D.~{Li}, Y.~{Han}, C.~{Wang}, G.~{Shi}, X.~{Wang}, X.~{Li}, and V.~C.~M.
  {Leung}, ``Deep reinforcement learning for cooperative edge caching in future
  mobile networks,'' in \emph{2019 IEEE Wireless Communications and Networking
  Conference (WCNC)}, 2019, pp. 1--6.

\bibitem{chien2020fgcs}
W.-C. Chien, H.-Y. Weng, and C.-F. Lai, ``Q-learning based collaborative cache
  allocation in mobile edge computing,'' \emph{Future Generation Computer
  Systems}, vol. 102, pp. 603 -- 610, 2020.

\bibitem{sung2016icmla}
J.~{Sung}, K.~{Kim}, J.~{Kim}, and J.~K. {Rhee}, ``Efficient content
  replacement in wireless content delivery network with cooperative caching,''
  in \emph{2016 15th IEEE International Conference on Machine Learning and
  Applications (ICMLA)}, 2016, pp. 547--552.

\bibitem{traverso2013sigcom}
S.~Traverso, M.~Ahmed, M.~Garetto, P.~Giaccone, E.~Leonardi, and S.~Niccolini,
  ``Temporal locality in today's content caching: Why it matters and how to
  model it,'' \emph{SIGCOMM Comput. Commun. Rev.}, vol.~43, no.~5, pp. 5--12,
  Nov. 2013.

\bibitem{gu2014icc}
J.~{Gu}, W.~{Wang}, A.~{Huang}, H.~{Shan}, and Z.~{Zhang}, ``Distributed cache
  replacement for caching-enable base stations in cellular networks,'' in
  \emph{2014 IEEE International Conference on Communications (ICC)}, 2014, pp.
  2648--2653.

\bibitem{tang2020tii}
J.~{Tang}, H.~{Tang}, X.~{Zhang}, K.~{Cumanan}, G.~{Chen}, K.~{Wong}, and J.~A.
  {Chambers}, ``Energy minimization in d2d-assisted cache-enabled internet of
  things: A deep reinforcement learning approach,'' \emph{IEEE Transactions on
  Industrial Informatics}, vol.~16, no.~8, pp. 5412--5423, 2020.

\bibitem{yin2018gsip}
J.~{Yin}, L.~{Li}, Y.~{Xu}, W.~{Liang}, H.~{Zhang}, and Z.~{Han}, ``Joint
  content popularity prediction and content delivery policy for cache-enabled
  d2d networks: A deep reinforcement learning approach,'' in \emph{2018 IEEE
  Global Conference on Signal and Information Processing (GlobalSIP)}, 2018,
  pp. 609--613.

\bibitem{jiang2018infocom}
W.~{Jiang}, G.~{Feng}, S.~{Qin}, and T.~S.~P. {Yum}, ``Efficient d2d content
  caching using multi-agent reinforcement learning,'' in \emph{IEEE INFOCOM
  2018 - IEEE Conference on Computer Communications Workshops (INFOCOM
  WKSHPS)}, 2018, pp. 511--516.

\bibitem{zhang2020network}
R.~{Zhang}, F.~R. {Yu}, J.~{Liu}, R.~{Xie}, and T.~{Huang},
  ``Blockchain-incentivized d2d and mobile edge caching: A deep reinforcement
  learning approach,'' \emph{IEEE Network}, vol.~34, no.~4, pp. 150--157, 2020.

\bibitem{zhang2020twc}
R.~{Zhang}, F.~R. {Yu}, J.~{Liu}, T.~{Huang}, and Y.~{Liu}, ``Deep
  reinforcement learning (drl)-based device-to-device (d2d) caching with
  blockchain and mobile edge computing,'' \emph{IEEE Transactions on Wireless
  Communications}, vol.~19, no.~10, pp. 6469--6485, 2020.

\bibitem{he2018gc}
Y.~{He}, C.~{Liang}, F.~R. {Yu}, and V.~C.~M. {Leung}, ``Integrated computing,
  caching, and communication for trust-based social networks: A big data drl
  approach,'' in \emph{2018 IEEE Global Communications Conference (GLOBECOM)},
  2018, pp. 1--6.

\bibitem{he2018wc}
Y.~{He}, F.~R. {Yu}, N.~{Zhao}, and H.~{Yin}, ``Secure social networks in 5g
  systems with mobile edge computing, caching, and device-to-device
  communications,'' \emph{IEEE Wireless Communications}, vol.~25, no.~3, pp.
  103--109, 2018.

\bibitem{he2020tnse}
Y.~{He}, C.~{Liang}, F.~R. {Yu}, and Z.~{Han}, ``Trust-based social networks
  with computing, caching and communications: A deep reinforcement learning
  approach,'' \emph{IEEE Transactions on Network Science and Engineering},
  vol.~7, no.~1, pp. 66--79, 2020.

\bibitem{wang2019access}
D.~{Wang}, H.~{Qin}, B.~{Song}, X.~{Du}, and M.~{Guizani}, ``Resource
  allocation in information-centric wireless networking with d2d-enabled mec: A
  deep reinforcement learning approach,'' \emph{IEEE Access}, vol.~7, pp.
  114\,935--114\,944, 2019.

\bibitem{sun2018jnca}
S.~Sun, M.~Liu, Z.~Jiao, X.~Pang, and S.~Chen, ``User-centric content sharing
  via cache-enabled device-to-device communication,'' \emph{Journal of Network
  and Computer Applications}, vol. 115, pp. 103 -- 115, 2018.

\bibitem{penda2017gc}
D.~D. {Penda}, N.~{Nomikos}, T.~{Charalambous}, and M.~{Johansson}, ``Minimum
  power scheduling under rician fading in full-duplex relay-assisted d2d
  communication,'' in \emph{2017 IEEE Globecom Workshops (GC Wkshps)}, 2017,
  pp. 1--6.

\bibitem{chen2017cl}
Z.~{Chen}, N.~{Pappas}, and M.~{Kountouris}, ``Probabilistic caching in
  wireless d2d networks: Cache hit optimal versus throughput optimal,''
  \emph{IEEE Communications Letters}, vol.~21, no.~3, pp. 584--587, 2017.

\bibitem{lin2019vtm}
P.~{Lin}, K.~S. {Khan}, Q.~{Song}, and A.~{Jamalipour}, ``Caching in
  heterogeneous ultradense 5g networks: A comprehensive cooperation approach,''
  \emph{IEEE Vehicular Technology Magazine}, vol.~14, no.~2, pp. 22--32, 2019.

\bibitem{chen2005ic}
T.~M. {Chen} and V.~{Venkataramanan}, ``Dempster-shafer theory for intrusion
  detection in ad hoc networks,'' \emph{IEEE Internet Computing}, vol.~9,
  no.~6, pp. 35--41, 2005.

\bibitem{bu2010milcom}
S.~{Bu}, F.~R. {Yu}, X.~L. {Peter}, H.~{Tang}, and P.~{Mason}, ``Distributed
  combined authentication and intrusion detection with data fusion in high
  security mobile ad-hoc networks,'' in \emph{2010 - MILCOM 2010 MILITARY
  COMMUNICATIONS CONFERENCE}, 2010, pp. 1080--1085.

\bibitem{su2015cm}
Z.~{Su} and Q.~{Xu}, ``Content distribution over content centric mobile social
  networks in 5g,'' \emph{IEEE Communications Magazine}, vol.~53, no.~6, pp.
  66--72, 2015.

\bibitem{shafiq2014acm}
M.~Z. Shafiq, A.~X. Liu, and A.~R. Khakpour, ``Revisiting caching in content
  delivery networks,'' vol.~42, no.~1, 2014.

\bibitem{jiang2016jsac}
J.~{Jiang}, S.~{Zhang}, B.~{Li}, and B.~{Li}, ``Maximized cellular traffic
  offloading via device-to-device content sharing,'' \emph{IEEE Journal on
  Selected Areas in Communications}, vol.~34, no.~1, pp. 82--91, 2016.

\bibitem{dai2019wc}
Y.~{Dai}, D.~{Xu}, S.~{Maharjan}, G.~{Qiao}, and Y.~{Zhang}, ``Artificial
  intelligence empowered edge computing and caching for internet of vehicles,''
  \emph{IEEE Wireless Communications}, vol.~26, no.~3, pp. 12--18, 2019.

\bibitem{dai2019wcsp}
Y.~{Dai}, D.~{Xu}, K.~{Zhang}, S.~{Maharjan}, and Y.~{Zhang}, ``Permissioned
  blockchain and deep reinforcement learning for content caching in vehicular
  edge computing and networks,'' in \emph{2019 11th International Conference on
  Wireless Communications and Signal Processing (WCSP)}, 2019, pp. 1--6.

\bibitem{dai2020tvt}
------, ``Deep reinforcement learning and permissioned blockchain for content
  caching in vehicular edge computing and networks,'' \emph{IEEE Transactions
  on Vehicular Technology}, vol.~69, no.~4, pp. 4312--4324, 2020.

\bibitem{dai2019iccc}
Y.~{Dai}, D.~{Xu}, Y.~{Lu}, S.~{Maharjan}, and Y.~{Zhang}, ``Deep reinforcement
  learning for edge caching and content delivery in internet of vehicles,'' in
  \emph{2019 IEEE/CIC International Conference on Communications in China
  (ICCC)}, 2019, pp. 134--139.

\bibitem{qiao2020iotj}
G.~{Qiao}, S.~{Leng}, S.~{Maharjan}, Y.~{Zhang}, and N.~{Ansari}, ``Deep
  reinforcement learning for cooperative content caching in vehicular edge
  computing and networks,'' \emph{IEEE Internet of Things Journal}, vol.~7,
  no.~1, pp. 247--257, 2020.

\bibitem{luo2020iotj}
Q.~{Luo}, C.~{Li}, T.~H. {Luan}, and W.~{Shi}, ``Collaborative data scheduling
  for vehicular edge computing via deep reinforcement learning,'' \emph{IEEE
  Internet of Things Journal}, vol.~7, no.~10, pp. 9637--9650, 2020.

\bibitem{he2020tvt}
Y.~{He}, N.~{Zhao}, and H.~{Yin}, ``Integrated networking, caching, and
  computing for connected vehicles: A deep reinforcement learning approach,''
  \emph{IEEE Transactions on Vehicular Technology}, vol.~67, no.~1, pp. 44--55,
  2018.

\bibitem{chen2020cc}
M.~Chen, T.~Wang, K.~Ota, M.~Dong, M.~Zhao, and A.~Liu, ``Intelligent resource
  allocation management for vehicles network: An a3c learning approach,''
  \emph{Computer Communications}, vol. 151, pp. 485 -- 494, 2020.

\bibitem{ning2020tits}
Z.~{Ning}, K.~{Zhang}, X.~{Wang}, M.~S. {Obaidat}, L.~{Guo}, X.~{Hu}, B.~{Hu},
  Y.~{Guo}, B.~{Sadoun}, and R.~Y.~K. {Kwok}, ``Joint computing and caching in
  5g-envisioned internet of vehicles: A deep reinforcement learning-based
  traffic control system,'' \emph{IEEE Transactions on Intelligent
  Transportation Systems}, pp. 1--12, 2020.

\bibitem{tan2018tvt}
L.~T. {Tan} and R.~Q. {Hu}, ``Mobility-aware edge caching and computing in
  vehicle networks: A deep reinforcement learning,'' \emph{IEEE Transactions on
  Vehicular Technology}, vol.~67, no.~11, pp. 10\,190--10\,203, 2018.

\bibitem{tan2019tvt}
L.~T. {Tan}, R.~Q. {Hu}, and L.~{Hanzo}, ``Twin-timescale artificial
  intelligence aided mobility-aware edge caching and computing in vehicular
  networks,'' \emph{IEEE Transactions on Vehicular Technology}, vol.~68, no.~4,
  pp. 3086--3099, 2019.

\bibitem{dai2020pc}
H.~Dai, H.~Zhang, B.~Wang, and L.~Yang, ``The multi-objective deployment
  optimization of uav-mounted cache-enabled base stations,'' \emph{Physical
  Communication}, vol.~34, pp. 114 -- 120, 2019.

\bibitem{zhang2020tvt}
T.~{Zhang}, Z.~{Wang}, Y.~{Liu}, W.~{Xu}, and A.~{Nallanathan}, ``Caching
  placement and resource allocation for cache-enabling uav noma networks,''
  \emph{IEEE Transactions on Vehicular Technology}, vol.~69, no.~11, pp.
  12\,897--12\,911, 2020.

\bibitem{chen2019tcom}
M.~{Chen}, W.~{Saad}, and C.~{Yin}, ``Echo-liquid state deep learning for
  360$^o$ content transmission and caching in wireless vr networks with
  cellular-connected uavs,'' \emph{IEEE Transactions on Communications},
  vol.~67, no.~9, pp. 6386--6400, 2019.

\bibitem{bennis2010gc}
M.~{Bennis} and D.~{Niyato}, ``A q-learning based approach to interference
  avoidance in self-organized femtocell networks,'' in \emph{2010 IEEE Globecom
  Workshops}, 2010, pp. 706--710.

\bibitem{chen2018tcom}
M.~{Chen}, W.~{Saad}, and C.~{Yin}, ``Virtual reality over wireless networks:
  Quality-of-service model and learning-based resource management,'' \emph{IEEE
  Transactions on Communications}, vol.~66, no.~11, pp. 5621--5635, 2018.

\bibitem{chen2017gc}
------, ``Liquid state machine learning for resource allocation in a network of
  cache-enabled lte-u uavs,'' in \emph{GLOBECOM 2017 - 2017 IEEE Global
  Communications Conference}, 2017, pp. 1--6.

\bibitem{chen2019twc}
------, ``Liquid state machine learning for resource and cache management in
  lte-u unmanned aerial vehicle (uav) networks,'' \emph{IEEE Transactions on
  Wireless Communications}, vol.~18, no.~3, pp. 1504--1517, 2019.

\bibitem{alhilo2020tits}
A.~{Al-Hilo}, M.~{Samir}, C.~{Assi}, S.~{Sharafeddine}, and D.~{Ebrahimi},
  ``Uav-assisted content delivery in intelligent transportation systems-joint
  trajectory planning and cache management,'' \emph{IEEE Transactions on
  Intelligent Transportation Systems}, pp. 1--13, 2020.

\bibitem{wu2020wcmc}
C.~{Wu}, S.~{Shi}, S.~{Gu}, L.~{Zhang}, and X.~{Gu}, ``Deep reinforcement
  learning-based content placement and trajectory design in urban cache-enabled
  uav networks,'' \emph{Wireless Communications and Mobile Computing}, vol.
  2020, pp. 1--11, 2020.

\bibitem{maghsudi2016cm}
S.~{Maghsudi} and E.~{Hossain}, ``Multi-armed bandits with application to 5g
  small cells,'' \emph{IEEE Wireless Communications}, vol.~23, no.~3, pp.
  64--73, 2016.

\bibitem{nomikos2020mlsp}
N.~{Nomikos}, S.~{Talebi}, R.~{Wichman}, and T.~{Charalambous}, ``Bandit-based
  relay selection in cooperative networks over unknown stationary channels,''
  in \emph{2020 IEEE 30th International Workshop on Machine Learning for Signal
  Processing (MLSP)}, 2020, pp. 1--6.

\bibitem{ghoorchian2021tccn}
S.~{Ghoorchian} and S.~{Maghsudi}, ``Multi-armed bandit for energy-efficient
  and delay-sensitive edge computing in dynamic networks with uncertainty,''
  \emph{IEEE Transactions on Cognitive Communications and Networking}, vol.~7,
  no.~1, pp. 279--293, 2021.

\bibitem{nomikos2021icc}
N.~{Nomikos}, T.~{Charalambous}, and R.~{Wichman}, ``Bandit-based power control
  in full-duplex cooperative relay networks, icc 2021, ieee international
  conference on communications,'' in \emph{ICC 2021 - 2019 IEEE International
  Conference on Communications (ICC)}, 2021, pp. 1--6.

\bibitem{petropulu2020icassp}
A.~Dimas, K.~Diamantaras, and A.~P. Petropulu, ``Q-learning based predictive
  relay selection for optimal relay beamforming,'' in \emph{ICASSP 2020 - 2020
  IEEE International Conference on Acoustics, Speech and Signal Processing
  (ICASSP)}, 2020, pp. 5030--5034.

\bibitem{petropulu2020tcom}
W.~Xia, G.~Zheng, Y.~Zhu, J.~Zhang, J.~Wang, and A.~P. Petropulu, ``A deep
  learning framework for optimization of miso downlink beamforming,''
  \emph{IEEE Transactions on Communications}, vol.~68, no.~3, pp. 1866--1880,
  2020.

\bibitem{psomas2020cl}
C.~Psomas and I.~Krikidis, ``Wireless powered mobile edge computing: Offloading
  or local computation?'' \emph{IEEE Communications Letters}, vol.~24, no.~11,
  pp. 2642--2646, 2020.

\bibitem{nomikos2018tvt}
N.~{Nomikos}, T.~{Charalambous}, D.~{Vouyioukas}, R.~{Wichman}, and G.~K.
  {Karagiannidis}, ``Power adaptation in buffer-aided full-duplex relay
  networks with statistical csi,'' \emph{IEEE Transactions on Vehicular
  Technology}, vol.~67, no.~8, pp. 7846--7850, 2018.

\bibitem{kim2019adhoc}
T.~Charalambous, S.~M. Kim, N.~Nomikos, M.~Bengtsson, and M.~Johansson,
  ``Relay-pair selection in buffer-aided successive opportunistic relaying
  using a multi-antenna source,'' \emph{Ad Hoc Networks}, vol.~84, pp. 29--41,
  2019.

\bibitem{simoni2016twc}
R.~{Simoni}, V.~{Jamali}, N.~{Zlatanov}, R.~{Schober}, L.~{Pierucci}, and
  R.~{Fantacci}, ``Buffer-aided diamond relay network with block fading and
  inter-relay interference,'' \emph{IEEE Transactions on Wireless
  Communications}, vol.~15, no.~11, pp. 7357--7372, 2016.

\bibitem{zhang2015cm}
H.~{Zhang}, Y.~{Li}, D.~{Jin}, M.~M. {Hassan}, A.~{Alelaiwi}, and S.~{Chen},
  ``Buffer-aided device-to-device communication: opportunities and
  challenges,'' \emph{IEEE Communications Magazine}, vol.~53, no.~12, pp.
  67--74, 2015.

\bibitem{ding2018tcom}
Z.~{Ding}, P.~{Fan}, G.~K. {Karagiannidis}, R.~{Schober}, and H.~V. {Poor},
  ``Noma assisted wireless caching: Strategies and performance analysis,''
  \emph{IEEE Transactions on Communications}, vol.~66, no.~10, pp. 4854--4876,
  2018.

\bibitem{xiang2019jstsp}
L.~{Xiang}, D.~W.~K. {Ng}, X.~{Ge}, Z.~{Ding}, V.~W.~S. {Wong}, and
  R.~{Schober}, ``Cache-aided non-orthogonal multiple access: The two-user
  case,'' \emph{IEEE Journal of Selected Topics in Signal Processing}, vol.~13,
  no.~3, pp. 436--451, 2019.

\bibitem{pei2020tvt}
X.~{Pei}, H.~{Yu}, Y.~{Chen}, M.~{Wen}, and G.~{Chen}, ``Hybrid
  multicast/unicast design in {NOMA}-based vehicular caching system,''
  \emph{IEEE Transactions on Vehicular Technology}, vol.~69, no.~12, pp.
  16\,304--16\,308, 2020.

\bibitem{luo2017cl}
S.~{Luo} and K.~C. {Teh}, ``Adaptive transmission for cooperative noma system
  with buffer-aided relaying,'' \emph{IEEE Communications Letters}, vol.~21,
  no.~4, pp. 937--940, 2017.

\bibitem{nomikos2020tvt}
N.~{Nomikos}, T.~{Charalambous}, D.~{Vouyioukas}, R.~{Wichman}, and G.~K.
  {Karagiannidis}, ``Integrating broadcasting and noma in full-duplex
  buffer-aided opportunistic relay networks,'' \emph{IEEE Transactions on
  Vehicular Technology}, vol.~69, no.~8, pp. 9157--9162, 2020.

\bibitem{xu2020iotj}
P.~{Xu}, Y.~{Wang}, G.~{Chen}, G.~{Pan}, and Z.~{Ding}, ``Design and evaluation
  of buffer-aided cooperative noma with direct transmission in iot,''
  \emph{IEEE Internet of Things Journal}, pp. 1--1, 2020.

\bibitem{li2020tcom}
J.~{Li}, X.~{Lei}, P.~D. {Diamantoulakis}, F.~{Zhou}, P.~{Sarigiannidis}, and
  G.~K. {Karagiannidis}, ``Resource allocation in buffer-aided cooperative
  non-orthogonal multiple access systems,'' \emph{IEEE Transactions on
  Communications}, vol.~68, no.~12, pp. 7429--7445, 2020.

\bibitem{Zhao:2018}
W.~Zhao, Z.~Chen, K.~Li, N.~Liu, B.~Xia, and L.~Luo, ``Caching-aided physical
  layer security in wireless cache-enabled heterogeneous networks,'' \emph{IEEE
  Access}, vol.~6, pp. 68\,920--68\,931, 2018.

\bibitem{ZhengTCOM:2018}
T.-X. Zheng, H.-M. Wang, and J.~Yuan, ``Secure and energy-efficient
  transmissions in cache-enabled heterogeneous cellular networks: Performance
  analysis and optimization,'' \emph{IEEE Transactions on Communications},
  vol.~66, no.~11, pp. 5554--5567, 2018.

\bibitem{Petropulu:2010}
L.~Dong, Z.~Han, A.~P. Petropulu, and H.~V. Poor, ``Improving wireless physical
  layer security via cooperating relays,'' \emph{IEEE Transactions on Signal
  Processing}, vol.~58, no.~3, pp. 1875--1888, 2010.

\bibitem{petropulu2020cl}
A.~Garnaev, A.~Petropulu, W.~Trappe, and H.~V. Poor, ``A multi-jammer game with
  latency as the user's communication utility,'' \emph{IEEE Communications
  Letters}, vol.~24, no.~9, pp. 1899--1903, 2020.

\end{thebibliography}
\let\addcontentsline\oldaddcontentsline% Restore \addcontentsline

\end{document}